\documentclass[fleqn,usenatbib]{aa}
\usepackage{txfonts}
\usepackage{graphicx}
\usepackage{natbib}
\usepackage{longtable}
\usepackage{subeqnarray}
\usepackage{cases}

\usepackage[colorlinks=true,citecolor=blue]{hyperref}

\begin{document}
\title{Kinetic temperature of massive star-forming molecular clumps measured with formaldehyde}
\subtitle{V. The massive filament DR21}

\author{{X. Zhao\inst{\ref{inst1},\ref{inst2}}}
\and X. D. Tang\inst{\ref{inst1},\ref{inst2},\ref{inst3},\ref{inst4}}
\and C. Henkel\inst{\ref{inst5},\ref{inst1}}
\and Y. Gong\inst{\ref{inst5},\ref{inst6}}
\and Y. Lin\inst{\ref{inst5},\ref{inst7}}
\and D. L. Li\inst{\ref{inst1},\ref{inst2},\ref{inst3},\ref{inst4}}
\and Y. X. He\inst{\ref{inst1},\ref{inst2},\ref{inst3},\ref{inst4}}
\and Y. P. Ao\inst{\ref{inst6}}
\and X. Lu\inst{\ref{inst8}}
\and T. Liu\inst{\ref{inst8}}
\and Y. Sun\inst{\ref{inst6}}
\and K. Wang\inst{\ref{inst9}}
\and X. P. Chen\inst{\ref{inst6}}
\and J. Esimbek\inst{\ref{inst1},\ref{inst3},\ref{inst4}}
\and J. J. Zhou\inst{\ref{inst1},\ref{inst3},\ref{inst4}}
\and J. W. Wu\inst{\ref{inst10},\ref{inst2}}
\and J. J. Qiu\inst{\ref{inst11}}
\and X. W. Zheng\inst{\ref{inst12}}
\and J. S. Li\inst{\ref{inst1},\ref{inst2}}
\and C. S. Luo\inst{\ref{inst1},\ref{inst2}}
\and Q. Zhao\inst{\ref{inst1},\ref{inst2}}
}


\titlerunning{Kinetic temperatures in DR21 filament}
\authorrunning{Zhao et al.}

\institute{
Xinjiang Astronomical Observatory, Chinese Academy of Sciences, 830011 Urumqi, PR China \label{inst1}\\
\email{tangxindi@xao.ac.cn}
\and University of Chinese Academy of Sciences, 100080 Beijing, PR China \label{inst2}
\and Key Laboratory of Radio Astronomy, Chinese Academy of Sciences, 830011 Urumqi, PR China \label{inst3}
\and Xinjiang Key Laboratory of Radio Astrophysics, Urumqi 830011, PR China \label{inst4}
\and Max-Planck-Institut f\"{u}r Radioastronomie, Auf dem H\"{u}gel 69, 53121 Bonn, Germany \label{inst5}
\and Purple Mountain Observatory, Chinese Academy of Sciences, Nanjing 210008, PR China \label{inst6}
\and Max-Planck-Institut f\"{u}r Extraterrestrische Physik, Giessenbachstr. 1, 85748 Garching bei M\"{u}nchen, Germany \label{inst7}
\and Shanghai Astronomical Observatory, Chinese Academy of Sciences, 80 Nandan Road, Shanghai 200030, PR China \label{inst8}
\and Kavli Institute for Astronomy and Astrophysics, Peking University, Beijing 100871, PR China \label{inst9}
\and National Astronomical Observatories, Chinese Academy of Sciences, Beijing 100101, PR China \label{inst10}
\and School of Mathmatics and Physics, Jinggangshan University, 343009 Ji’an, PR China \label{inst11}
\and School of Astronomy and Space Science, Nanjing University, 210093 Nanjing, PR China \label{inst12}
}

\abstract
{The kinetic temperature structure of the massive filament DR21 within the Cygnus X molecular cloud complex has been mapped using the
IRAM\,30\,m telescope. This mapping employed the para-H$_2$CO triplet ($J_{\rm K_aK_c}$\,=\,3$_{03}$--2$_{02}$, 3$_{22}$--2$_{21}$,
and 3$_{21}$--2$_{20}$) on a scale of $\sim$0.1\,pc.
By modeling the averaged line ratios of para-H$_{2}$CO\,3$_{22}$--2$_{21}$/3$_{03}$--2$_{02}$ and 3$_{21}$--2$_{20}$/3$_{03}$--2$_{02}$
with RADEX under non-LTE assumptions, the kinetic temperature of the dense gas was derived, which ranges from 24 to 114\,K,
with an average temperature of 48.3\,$\pm$\,0.5\,K at a density of $n$(H$_{2}$)\,=\,10$^{5}$\,cm$^{-3}$.
In comparison to temperature measurements using NH$_3$\,(1,1)/(2,2) and FIR wavelengths,
the para-H$_2$CO\,(3--2) lines reveal significantly higher temperatures.
The dense clumps in various regions appear to correlate with the notable kinetic temperature ($T_{\rm kin}$\,$\gtrsim$\,50\,K) of
the dense gas traced by H$_2$CO. Conversely, the outskirts of the DR21 filament display lower temperature distributions ($T_{\rm kin}$\,$<$\,50\,K).
Among the four dense cores (N44, N46, N48, and N54), temperature gradients are observed on a scale of $\sim$0.1-0.3\,pc.
This suggests that the warm dense gas traced by H$_2$CO is influenced by internal star formation activity. With the exception of
the dense core N54, the temperature profiles of these cores were fitted with power-law indices ranging from $-$0.3 to $-$0.5,
with a mean value of approximately $-$0.4. This indicates that the warm dense gas probed by H$_2$CO is heated by radiation emitted
from internally embedded protostar(s) and/or clusters. While there is no direct evidence supporting the idea that the dense gas is
heated by shocks resulting from a past explosive event in the DR21 region on a scale of $\sim$0.1\,pc, our measurements of H$_2$CO
towards the DR21W1 region provide compelling evidence that the dense gas in this specific area is indeed heated by shocks originating
from the western DR21 flow. Higher temperatures as traced by H$_2$CO appear to be associated with turbulence on a scale of $\sim$0.1\,pc.
The physical parameters of the dense gas as determined from H$_2$CO lines in the DR21 filament exhibit a remarkable similarity to
the results obtained in OMC-1 and N113 albeit on a scale of approximately 0.1-0.4\,pc. This may imply that the physical mechanisms
governing the dynamics and thermodynamics of dense gas traced by H$_{2}$CO in diverse star formation regions may be dominated by common
underlying principles despite variations in specific environmental conditions.}

 \maketitle

\section{Introduction}
\label{sect:Introduction}
Increasing evidence indicates that the physical properties of the interstellar medium (ISM) have an impact on the star
formation rate, the spatial distribution and essential properties of the next generation of stars, such as
the elemental composition and initial mass function (e.g.,
\citealt{Paumard2006,Kennicutt1998a,Kennicutt1998b,Klessen2007,Papadopoulos2011,Zhang2018,Tang2019}).
High-mass stars exert significant influence on their surroundings and subsequent star formation through
their feedback such as outflows, winds, and UV radiation. However, the specifics of the high-mass star
formation process and how their feedback affects the initial conditions of high-mass stars during
formation are not yet fully understood. To investigate the intricacies or star formation within molecular gas,
it is crucial to obtain accurate measurements of its physical properties
such as kinetic temperature and volume density.

\subsection{DR21 Filament}
\label{sect:DR21-filament}
The Cygnus X molecular cloud complex provides a large sample of star formation regions that exhibit diverse physical conditions
and evolutionary stages, making it an ideal laboratory for investigating the process of star formation
(e.g., \citealt{Reipurth2008,Schneider2010}). The DR21 filament at a distance of $\sim$1.5\,kpc \citep{Rygl2012} is the densest
and most massive region in the entire Cygnus X molecular complex, which is one of the most active star-forming regions within
2\,kpc from the Sun (e.g., \citealt{Schneider2006,Motte2007,Davis2007,Kumar2007,Beerer2010,Csengeri2011}).
The DR21 filament is an elongated and dense molecular ridge that spans $\sim$4\,pc in length, extending in a North-South direction.
It hosts a mass of $\sim$10$^4$\,M$_{\odot}$, which is typically defined by high column densities,
i.e., $N$(H$_2$)\,>\,10$^{23}$\,cm$^{-2}$ \citep{Schneider2010,Hennemann2012}.
The DR21 filament encompasses three prominent regions, namely DR21, DR21(OH), and W75S.
Additionally, the DR21W1 region situated to the west of the DR21 region is associated with the western DR21 flow
(e.g., \citealt{Jaffe1989,Lane1990,Davis2007,Zapata2013}). The two well-known star-forming regions DR21 and DR21(OH) are located
in the southern and central parts of the DR21 filament, respectively. 
DR21 is a relatively more evolved massive star-forming region, which harbors a group of luminous H\,{\scriptsize II} regions and
massive protostars associated with very energetic outflows (e.g., \citealt{Harris1973,Garden1991a,Russell1992,Kogan1998,Cyganowski2003,Rygl2012}).
The observed outflows in DR21 exhibit a bipolar structure, characterized by its most prominent molecular lobes extending in a northeast-southwest
direction (e.g., \citealt{Davis2007,Schneider2010,Skretas2023}). DR21(OH) is an extremely massive star-forming region with intensive maser activity
(e.g., \citealt{Genzel1977,Norris1982,Batrla1988,Kogan1998,Harvey-Smith2008,Schneider2010}). The W75S region, which contains several
far-infrared (FIR) sources, is positioned to the north of the DR21 filament \citep{Harvey1986,Wilson1990,Davis2007,Kumar2007,Reipurth2008}.

Extensive investigations of the DR21 filament have been conducted across various wavelengths, including radio, millimeter, and infrared wavelengths.
The region displays a remarkable complexity, with numerous infrared filaments oriented perpendicularly to the ridge, indicating potential
sites of star formation as evidenced by observations from the {\it Spitzer} and {\it Herschel} telescopes \citep{Marston2004,Kumar2007,Beerer2010,Hennemann2012}.
Dust continuum studies reveal a chain of dusty, dense cores along the ridge of the DR21 filament \citep{Chandler1993,Vallee2006,Motte2007,Cao2019}.
A large number of molecular line observations have been performed, such as in CO, CN, CS, HCN, HCO$^{+}$, CH$_{3}$OH, H$_{2}$CO,
NH$_3$, and N$_{2}$H$^{+}$ (e.g., \citealt{Wilson1990,Vallee2006,Schneider2006,Schneider2010,Schneider2016,Zapata2012,Dobashi2019,Keown2019,Cao2022,Bonne2023,Gong2023b}),
which reveal the dynamics and evolution of the dense molecular structure of the DR21 filament on large scales.

The gas temperature structure of the DR21 filament has been mapped in NH$_3$\,(1,1) and (2,2) with the GBT
(beam size $\sim$32$''$; \citealt{Keown2019}). The typical kinetic temperature of the dense gas, derived from
NH$_3$\,(2,2)/(1,1) line ratios, fall within the range of 20–30\,K. In specific regions, the typical gas kinetic temperatures are
around 20\,K in the W75S region, approximately 22--64\,K in the DR21(OH) region, and about 35--60\,K in the DR21W1 region.
\cite{Mangum1992} presented high-resolution imaging of the NH$_3$\,(1,1) and (2,2) lines within the DR21(OH) region, utilizing the Very
Large Array (VLA) with a beam size of $\sim$4$''$. By also analyzing the line ratios of NH$_3$\,(2,2)/(1,1) within dense ammonia cores,
they derived the gas kinetic temperatures. These temperatures range from 20 to $>$80\,K, with a typical value of approximately 30\,K.
The dust temperature structure of the DR21 filament has been derived from spectral energy distribution (SED) fitting to {\it Herschel} continuum
data spanning 70--500\,$\mu$m (beam size $\sim$40$''$; \citealt{Hennemann2012,Cao2019,Cao2022}). The typical dust temperature was found to be between 15 and 31\,K,
with specific regions exhibiting typical values of 20, 21, 22, 20, 31, 23, 21, 18, and 16\,K for W75S\,FIR\,3, W75S\,FIR\,2, W75S\,FIR\,1, DR21(OH)N,
DR21(OH), DR21(OH)S, DR21, DR21W1, and DR21S, respectively.
These studies provide valuable insights into the gas and dust temperature variations within the DR21 filament.
Nevertheless, it is worth noting that the NH$_3$\,(1,1) and (2,2) lines exhibit critical densities on the order of
$n_{\rm crit}$(NH$_{3}$)\,$\sim$\,10$^{3}$\,cm$^{-3}$ \citep{Shirley2015}. This characteristic suggests that these particular spectral
lines possess a sensitivity suitable for probing the moderate-density gas within molecular clouds.
A more comprehensive database of gas temperature which is related to star formation ($n$(H$_{2}$)>10$^{4}$\,cm$^{-3}$) in the DR21 filament,
with high resolution at a sub-pc scale, is still lacking. Hence, further research and high-resolution observations of the DR21 filament
are necessary to achieve a comprehensive characterization of the thermal properties of the dense gas.

\subsection{H$_2$CO as a Temperature Tracer}
Formaldehyde (H$_2$CO), a slightly asymmetric rotor, is ubiquitous in interstellar clouds
(e.g., \citealt{Downes1980,Hurt1996,Muhle2007,Ao2013,Tang2013,Tang2014,Guo2016,Yan2019,Brinkmann2020}), exhibits a large number of transitions
from centimeter to millimeter wavelengths and is thought to be sensitive to probe physical conditions in various molecular clouds
(e.g., \citealt{Henkel1980,Henkel1983,Mangum1993a,Mangum2008,Mangum2013a,Mangum2019,Ginsburg2011,Ginsburg2016,Ao2013,Tang2017a,Tang2017b,Tang2018a,Tang2018b,Tang2021,Immer2016}).
Previous observations of H$_2$CO indicate that variations in the fractional abundance of H$_{2}$CO only rarely exceed
one order of magnitude at various stages of star formation
(e.g., \citealt{Mangum1990,Mangum1993b,Caselli1993,Johnstone2003,Gerner2014,Tang2017a,Tang2017b,Tang2018b,Liu2020}).
It is possibly the best of the very few molecular thermometers that are available for the analysis of dense molecular gas.

Since the relative populations of the $K_{\rm a}$ ladders of H$_2$CO are predominantly governed
by collisions, ratios of H$_2$CO line fluxes involving different $K_{\rm a}$ ladders can act as good indicators of kinetic temperature,
such as para-H$_{2}$CO\,$J_{\rm K_aK_c}$\,=\,3$_{22}$--2$_{21}$/3$_{03}$--2$_{02}$, 4$_{22}$--3$_{21}$/4$_{04}$--3$_{03}$,
and 5$_{23}$--4$_{22}$/5$_{05}$--4$_{04}$  \citep{Mangum1993a,Tang2018b,Mazumdar2022,Kahle2023}. These H$_2$CO line ratios can accurately trace dense
molecular gas as thermometers and have been used to measure the temperature of dense gas in Galactic star
formation regions (e.g., \citealt{Mangum1993a,Hurt1996,Mangum1999,Watanabe2008,Nagy2012,Lindberg2015,Tang2017a,Tang2018a,Tang2018b,Gieser2021,Izumi2023}),
Galactic center clouds (\citealt{Qin2008,Ao2013,Johnston2014,Ginsburg2016,Immer2016,Lu2017}),
and even external galaxies (\citealt{Muhle2007,Tang2017b,Tang2021,Mangum2019}).

To understand the feedback from the high-mass star-forming process, it is critical to investigate the correlation between
the gas temperature and star formation activities in the molecular clouds.
The main goal of this paper is to map the kinetic temperature structure
of the massive DR21 filament in the north of the Cygnus X molecular cloud complex, making use of the
para-H$_2$CO triplet ($J_{\rm K_aK_c}$\,=\,3$_{03}$--2$_{02}$, 3$_{22}$--2$_{21}$,
and 3$_{21}$--2$_{20}$, hereafter H$_{2}$CO) with a critical density on the order of
$n_{\rm crit}$(H$_2$CO\,3$_{03}$--2$_{02}$)\,$\sim$\,a few\,10$^{5}$\,cm$^{-3}$ \citep{Shirley2015}
and to investigate the gas heating mechanism
affecting the dense gas in the processes of star formation.
In Sects.\,\ref{sect:Observations-and-data-reduction} and \ref{sect:Results},
we introduce our observations of the H$_2$CO triplet,
data reduction, and results. We discuss the resulting
kinetic temperatures derived from H$_2$CO in Sect.\,\ref{sect:discussion}.
Our main conclusions are summarized in Sect.\,\ref{Sec:Sumamry}.
This paper is part of the "Kinetic temperature of massive
star-forming molecular clumps measured with formaldehyde" series of
studies exploiting H$_2$CO as a probe of gas conditions in a variety
of Galactic and extragalactic sources.

\section{Observations and Data Reduction}
\label{sect:Observations-and-data-reduction}
Our data presented here were observed with the IRAM\,30\,m telescope\footnote{\tiny Based on observations obtained with the IRAM\,30\,m telescope.
IRAM is supported by INSU/CNRS (France), MPG (Germany), and IGN (Spain).} in 2017 May and 2018 October using the HERA receiver \citep{Schuster2004}.
The H$_{2}$CO ($J_{\rm K_aK_c}$\,=\,3$_{03}$--2$_{02}$, 3$_{22}$--2$_{21}$ and 3$_{21}$--2$_{20}$) transitions have rest frequencies of 218.222, 218.475,
and 218.760\,GHz, respectively, and could be measured simultaneously. At $\sim$218\,GHz, the beam size is $\sim$\,12$''$
($\sim$\,0.09\,pc at 1.5\,kpc distance; \citealt{Rygl2012}). The main beam efficiency and the forward
efficiency\footnote{\tiny https://publicwiki.iram.es/Iram30mEfficiencies} are 0.60 and 0.94, respectively.
The backend spectrometer has a bandwidth of $\sim$1\,GHz and 5377 spectral channels, providing a channel width of $\sim$0.27\,km\,s$^{-1}$.
The central frequency was set to 218.475\,GHz. The on-the-fly observing mode was used to measure four $\sim$3$\times$3\,arcmin$^2$ maps with steps
of 3.5$''$ in both right ascension and declination. The surveyed area of the DR21 filament is $\sim$4$\times$10\,arcmin$^2$ ($\sim$1.7$\times$4.4\,pc$^{2}$),
centered on $\alpha_{2000}$\,=\,20$^h$39$^m$01\hbox{$\,.\!\!^s$}0 and $\delta_{2000}$\,=\,42$\degr$22$'$49\hbox{$\,.\!\!^{\prime\prime}$}8.

Data reduction for spectral lines and maps was performed with GILDAS\footnote{\tiny http://www.iram.fr/IRAMFR/GILDAS}.
To improve the signal-to-noise ratio (S/N) in individual channels, we smoothed contiguous channels to a velocity resolution
$\sim$0.8\,km\,s$^{-1}$ and the spectra were resampled in steps of $\sim$6$''$.
In total, we acquired 3200 positions corresponding to 3200 spectra for each transition.
Nearly 55\% of all positions were detected in H$_{2}$CO\,3$_{03}$--2$_{02}$ and $\sim$23\% were also detected in
H$_{2}$CO\,3$_{22}$--2$_{21}$ and 3$_{21}$--2$_{20}$, with S/Ns of $\gtrsim$3$\sigma$.
We used Gaussian profiles to fit all H$_{2}$CO spectra with S/Ns above 3$\sigma$.
A typical rms noise level (1$\sigma$) is $\sim$0.19\,K ($T_{\rm mb}$ scale) at a channel width of 0.8\,km\,s$^{-1}$.

Generally, H$_{2}$CO\,3$_{22}$--2$_{21}$ and 3$_{21}$--2$_{20}$ show weaker emissions than H$_{2}$CO\,3$_{03}$--2$_{02}$.
The H$_{2}$CO\,3$_{22}$--2$_{21}$ and 3$_{21}$--2$_{20}$ transitions have the same $K_{\rm a}$ ladder
and almost the same energy of the upper levels, $E_{\rm u}$\,$\simeq$\,68\,K, above the ground state.
They also show similar distributions and line profiles (brightness temperature, linewidth, and velocity in our observations)
in star formation regions (e.g., \citealt{Mangum1993a,Tang2017a,Tang2017b,Tang2018a,Tang2018b,Tang2021}).
Furthermore, H$_{2}$CO 3$_{22}$--2$_{21}$/3$_{03}$--2$_{02}$ and 3$_{21}$--2$_{20}$/3$_{03}$--2$_{02}$ line ratios exhibit similar
behaviors to trace the kinetic temperature of dense gas (e.g., \citealt{Mangum1993a,Tang2017a}).
To further enhance the S/Ns of H$_{2}$CO\,3$_{22}$--2$_{21}$ and 3$_{21}$--2$_{20}$, we combined the line intensity
of two transitions channel by channel.

In the vicinity of DR21(OH), the H$_{2}$CO\,3$_{03}$--2$_{02}$ spectral lines possess two distinct velocity components.
However, identifying these two distinct velocity components becomes difficult for the weaker H$_{2}$CO\,3$_{22}$--2$_{21}$
and 3$_{21}$--2$_{20}$ lines. To address this, we employed a Gaussian method to fit these $\sim$70 positions that clearly
exhibit two velocity components. The two velocity components measured using H$_{2}$CO\,3$_{03}$--2$_{02}$ are approximately
at local standard of rest velocities ($V_{\rm LSR}$) of $-$4 and $-$2\,km\,s$^{-1}$,
consistent with previous findings from observations of HCO$^+$, H$^{13}$CO$^+$, and N$_2$H$^+$ \citep{Schneider2010,Cao2022,Bonne2023}.
High-resolution observations revealed that these two velocity components were attributed to two closely aligned fibers in the
DR21(OH)/W75S region \citep{Cao2022}. Subsequently, we combined the data from these two velocity components based on the integrated
intensity ratio between each pair of velocity component lines. Hence, the gas temperatures of these positions in the two velocity
components were not determined individually.

\section{Results}
\label{sect:Results}
\subsection{Overview}
\label{sec:Overview}
The velocity-integrated intensity distributions of the H$_2$CO triplet and the combined H$_{2}$CO\,3$_{22}$--2$_{21}$ and 3$_{21}$--2$_{20}$
lines in the DR21 filament ($V_{\rm LSR}$\,=\,$-$10 to 4\,km\,s$^{-1}$) are shown in Fig.\,\ref{fig:H2CO-intensity-maps}.
The intensity-weighted velocity field (moment 1), linewidth (moment 2), and channel maps of H$_{2}$CO\,3$_{03}$--2$_{02}$
are shown in Figs.\,\ref{fig:H2CO-moment1&2} and \ref{fig:H2CO-channel}. The observed H$_2$CO spectra of sixteen dense cores identified
with 1.2\,mm continuum emission by \cite{Motte2007} are shown in Fig.\,\ref{fig:H2CO-core}. The locations of these dense cores in
the DR21 filament are indicated in Fig.\,\ref{fig:H2CO-ratio} (left panel) and are
listed in Table\,\ref{table:Clumps-Parameters}. The parameters of the Gaussian fits, which include velocity-integrated
intensity, $\int T_{\rm mb}$\,d$v$, local standard of rest velocity, $V_{\rm LSR}$, full width at half maximum (FWHM) line width,
and peak temperature ($T_{\rm mb}$) of the H$_2$CO spectra in these dense cores are listed in Table\,\ref{table:Clumps-Parameters}.
Three dense cores, namely N35, N45, and N52, situated at the periphery of the DR21 filament, exhibit weak emissions of H$_{2}$CO\,3$_{22}$--2$_{21}$ and 3$_{21}$--2$_{20}$ with S/Ns below 3$\sigma$, as depicted in Figs.\,\ref{fig:H2CO-intensity-maps} and \ref{fig:H2CO-ratio}. Consequently, these cores were omitted from Fig.\,\ref{fig:H2CO-core}.

\subsection{Distribution of H$_{2}$CO}
\label{sec:distribution-H2CO}
\begin{figure*}[t]
\centering
\includegraphics[width=1.0\textwidth,angle=0]{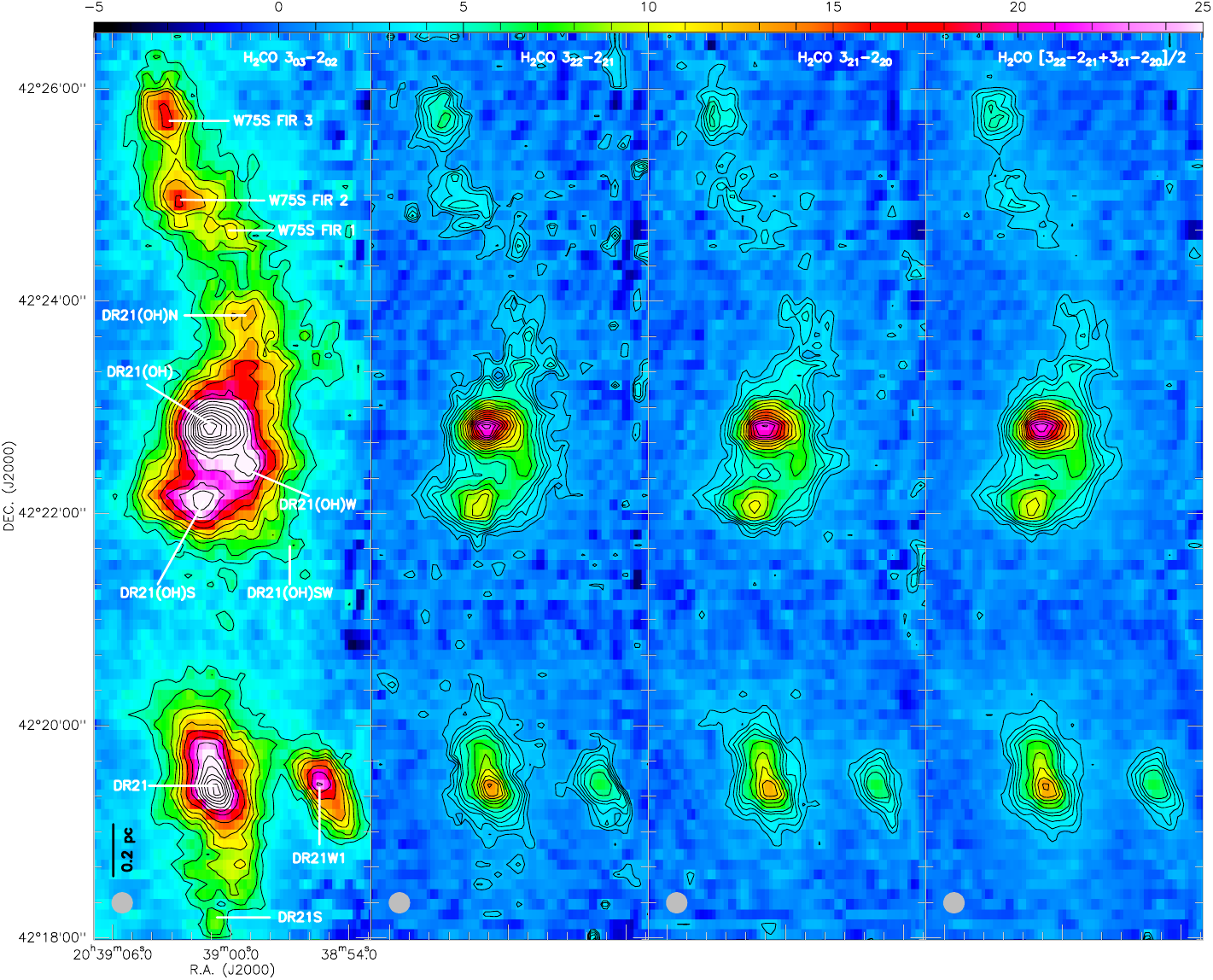}
\caption{Velocity-integrated intensity maps ($T_{\rm mb}$ scale; color bar in units of
K\,km\,s$^{-1}$) of H$_{2}$CO\,3$_{03}$--2$_{02}$ (\emph{left}),
3$_{22}$--2$_{21}$ (\emph{center-left}), 3$_{21}$--2$_{20}$ (\emph{center-right}),
and averaged 3$_{22}$--2$_{21}$ and 3$_{21}$--2$_{20}$ (\emph{right}),
integrated from $V_{\rm LSR}$\,=\,$-$10 to 4\,km\,s$^{-1}$ of the DR21 filament.
Contour levels are from 4.8 to 12\,K\,km\,s$^{-1}$ with steps of
1.8\,K\,km\,s$^{-1}$ and from 12 to 59\,K\,km\,s$^{-1}$ with steps
of 3.6\,K\,km\,s$^{-1}$ for H$_{2}$CO\,3$_{03}$--2$_{02}$,
and from 1.7 to 4.4\,K\,km\,s$^{-1}$ with steps of 0.66\,K\,km\,s$^{-1}$
and from 4.3 to 21.2\,K\,km\,s$^{-1}$ with steps of 1.3\,K\,km\,s$^{-1}$
for H$_{2}$CO\,3$_{22}$--2$_{21}$, 3$_{21}$--2$_{20}$, and combined 3$_{22}$--2$_{21}$ and 3$_{21}$--2$_{20}$.
The locations of source names are taken from \cite{Davis2007}, \cite{Reipurth2008}, and \cite{Ching2022}.
}
\label{fig:H2CO-intensity-maps}
\end{figure*}

The H$_{2}$CO\,3$_{03}$--2$_{02}$ line exhibits an extensive distribution and provides a clear picture of the dense ridge structure of the DR21 filament
(see the left panel of Fig.\,\ref{fig:H2CO-intensity-maps}). It illuminates notable dense molecular structures including the DR21, DR21(OH),
and W75S regions on a linear scale of $\sim$0.1\,pc. Obviously, the strongest emission of H$_{2}$CO is associated with the
two most active star-forming regions, DR21 and DR21(OH). These results are consistent with prior observational findings using other dense gas tracers,
such as CN, C$^{18}$O, CS, HCN, and HCO$^+$ (\citealt{Schneider2010,Dobashi2019}). The spatial distribution of H$_{2}$CO\,3$_{03}$--2$_{02}$
illustrates a close correspondence with the spatial distribution of the dust emission at wavelengths of 850\,$\mu {\rm m}$ and 1.2\,mm,
including the presence of dense ridge structures and dust emission peaks within the DR21 filament \citep{Davis2007,Motte2007,Ching2022},
which confirms previous observational results regarding massive star formation regions (e.g., \citealt{Tang2017a,Tang2018a,Tang2018b,Tang2021}).
In contrast, H$_{2}$CO\,3$_{22}$--2$_{21}$ and 3$_{21}$--2$_{20}$ indicate less extended distributions and are only detectable in
the densest areas of the DR21 filament, as shown in the rightmost three panels of Fig.\,\ref{fig:H2CO-intensity-maps}.

The intensity-weighted velocity  (moment 1) and channel maps of H$_{2}$CO\,(3$_{03}$--2$_{02}$) reveal the complex velocity field of the DR21 filament
(see the left panels of Figs.\,\ref{fig:H2CO-moment1&2} and \ref{fig:H2CO-channel}).
Notably, multiple velocity gradients traverse the primary DR21 filament as well as the DR21W1 region.
The W75S region shows an increasing gradient of velocity ($-$4.5 to $-$2.5\,km\,s$^{-1}$) that extends from the northeast to the southwest, while the DR21(OH) region
displays a decreasing northwest-southeast velocity gradient ($-$2.5 to $-$4.0\,km\,s$^{-1}$).
Utilizing high-resolution observations of H$^{13}$CO$^+$, N$_2$H$^+$, and NH$_2$D toward the W75S/DR21(OH) region (beam size $\sim$ 5$''$),
it has been observed that three nearly parallel fibers exist at distinct velocities, primarily aligned in a north-south orientation \citep{Cao2022}.
Such an alignment may give rise to the complex velocity field found within this region. Furthermore, the DR21 and DR21W1 regions exhibit a comparable
increasing northeast-to-southwest velocity gradient with the mean velocity ranging from $-$3.5 to $-$1.5\,km\,s$^{-1}$ and from $-$5.0 to $-$2.0\,km\,s$^{-1}$, respectively.
Our observed H$_{2}$CO results corroborate previous observational outcomes obtained from H$^{13}$CO$^+$, C$^{34}$S, and N$_2$H$^+$ \citep{Schneider2010}.

As shown in the right panel of Fig.\,\ref{fig:H2CO-moment1&2}, higher linewidths ($\gtrsim$3\,km\,s$^{-1}$)
are typically associated with dense clumps, such as DR21, DR21W1, DR21(OH), DR21(OH)S, and DR21(OH)W,
in the two most active star-forming regions DR21 and DR21(OH). Moderate linewidths
(2--3\,km\,s$^{-1}$) are distributed in the dense clumps of the W75S region, as well as around the dense clumps of DR21 and DR21(OH) regions,
while lower linewidths  ($<$2\,km\,s$^{-1}$) are located on the outskirts of the DR21 filament.
The fitted linewidth of H$_{2}$CO\,(3$_{03}$--2$_{02}$) ranges from 1.7 to 10.4\,km\,s$^{-1}$
with an average of $\sim$3.8$\pm$0.1\,km\,s$^{-1}$ (errors given here and elsewhere are standard deviations of the mean).

\subsection{Line Ratios of H$_{2}$CO}
\label{sec:H2CO-Ratio}
As mentioned in Sects.\,\ref{sect:Introduction} and \ref{sect:Observations-and-data-reduction},
H$_{2}$CO\,3$_{22}$--2$_{21}$/3$_{03}$--2$_{02}$ and 3$_{21}$--2$_{20}$/3$_{03}$--2$_{02}$ line ratios exhibit similar behaviors
in delineating the kinetic temperature of the dense gas (e.g., \citealt{Mangum1993a,Tang2017a}).
The present study employs the averaged ratio of H$_{2}$CO\,0.5$\times$[(3$_{22}$--2$_{21}$\,+\,3$_{21}$--2$_{20})$/3$_{03}$--2$_{02}$]
(obtained from the combined H$_{2}$CO\,3$_{22}$--2$_{21}$ and 3$_{21}$--2$_{20}$ to H$_{2}$CO\,3$_{03}$--2$_{02}$ ratio) as an indicator
of the kinetic temperature. Substantially, higher H$_{2}$CO line ratios typically correspond to higher kinetic
temperatures \citep{Ao2013,Ginsburg2016,Tang2018a,Tang2021}. Thus, the ratio maps may function as a proxy for the
relative kinetic temperature. The line ratio map of the DR21 filament is shown in the left panel of Fig.\,\ref{fig:H2CO-ratio}.
The line ratios are calculated by velocity-integrated intensities where the combined H$_{2}$CO\,3$_{22}$--2$_{21}$
and 3$_{21}$--2$_{20}$ lines are detected with S/Ns of $\gtrsim$\,3$\sigma$.
The H$_{2}$CO line ratios range from 0.08 to 0.42 with an average value of 0.22\,$\pm$\,0.01 in the DR21 filament.
Higher H$_{2}$CO line ratios (>0.35) have been observed in association with dense clumps, such as W75S\,FIR\,3, DR21(OH), DR21(OH)S, DR21, and DR21W1.
Moderate line ratios ranging from 0.2 to 0.35 are prevalent throughout the vicinity of these dense clumps. Notably, the outskirts of the DR21
and the area located between DR21(OH) and W75S exhibit relatively lower line ratios (<0.2).

\subsection{Kinetic Temperatures Derived from H$_{2}$CO Line Ratios}
\label{sec:Tk-H2CO}
As described in Sects.\,\ref{sect:Observations-and-data-reduction} and \ref{sec:H2CO-Ratio}, the determination of the kinetic temperature
involves calculating the averaged ratio of H$_{2}$CO\,0.5$\times$[(3$_{22}$--2$_{21}$\,+\,3$_{21}$--2$_{20})$/3$_{03}$--2$_{02}$].
To model the relationship between the gas kinetic temperature and the measured average of
H$_{2}$CO\,0.5$\times$[(3$_{22}$--2$_{21}$\,+\,3$_{21}$--2$_{20})$/3$_{03}$--2$_{02}$] ratios, we employed the
RADEX\footnote{\tiny http://var.sron.nl/radex/radex.php} non-LTE model \citep{van2007} with collision rates provided by \cite{Wiesenfeld2013}.
In Fig.\,\ref{fig:H2CO-radex}, we adopted a background temperature of 2.73\,K, an average measured linewidth of 3.8\,km\,s$^{-1}$,
and a column density of $N$(para-H$_{2}$CO)\,=\,6.6$\times$10$^{13}$\,cm$^{-2}$.
The impact of different column densities of H$_{2}$CO on the kinetic temperature is found to be minimal when all lines are optically thin,
as shown in Fig.\,3 in \cite{Tang2017b} or Fig.\,4 in \cite{Tang2018a}. Additionally, opacities within dense massive star-forming clumps have
little effect on the kinetic temperature \citep{Ginsburg2016,Immer2016,Tang2017b,Tang2018b}.
We assume that para-H$_{2}$CO\,(3--2) is optically thin in DR21 (see below for detailed arguments).
The measured H$_{2}$ number densities, $n$(H$_{2}$), of massive dense cores in DR21 range from
$\sim$\,10$^{4}$ to 10$^{6}$\,cm$^{-3}$ \citep{Schneider2006,Jakob2007,Cao2019,Cao2022}.
Previous observations have demonstrated that para-H$_{2}$CO\,(3--2) is particularly sensitive to the gas density at
$\sim$10$^{5}$\,cm$^{-3}$ \citep{Ginsburg2016,Immer2016,Tang2017b,Tang2021}. Therefore, we adopt an average spatial gas density of
10$^{5}$\,cm$^{-3}$ for the DR21 filament. By utilizing the methodology outlined in Sect.\,3.1 of \cite{Tang2017b}, we can calculate the average
column density $N$(para-H$_{2}$CO) of the entire DR21 region based on the H$_{2}$CO\,(3$_{03}$--2$_{02}$) average brightness temperature.
The resulting value is 6.6$\times$10$^{13}$\,cm$^{-2}$ at a volume density of approximately 10$^{5}$\,cm$^{-3}$.
Previous observations of para-H$_{2}$CO\,(3--2 and 4--3) toward dense clumps representing various
evolutionary stages of high-mass star formation in the Galactic plane suggest an averaged fractional
abundance of para-H$_{2}$CO of $\sim$3.9$\times$10$^{-10}$ \citep{Tang2018b}. The average column density
of H$_2$, $N$(H$_2$), in the DR21 ridge, where the distribution of $N$(H$_2$) agrees well with that of
our H$_{2}$CO\,(3$_{03}$--2$_{02}$) lines, is $\sim$4.2$\times$10$^{23}$\,cm$^{-2}$ as determined from the
{\it Herschel} 70--500\,$\mu$m continuum data \citep{Hennemann2012}. Subsequently, we can compute a typical
fractional abundance of para-H$_{2}$CO of roughly 1.6$\times$10$^{-10}$ in the DR21 filament using the
above-mentioned average H$_{2}$CO column density of the entire DR21 region, which is in agreement with
previous observational results in high-mass star formation regions \citep{Tang2018b}. This supports the
reasonability of our RADEX calculations using the average column density $N$(para-H$_{2}$CO) in the DR21 region.

\begin{figure}[t]
\vspace*{0.2mm}
\begin{center}
\includegraphics[width=0.5\textwidth]{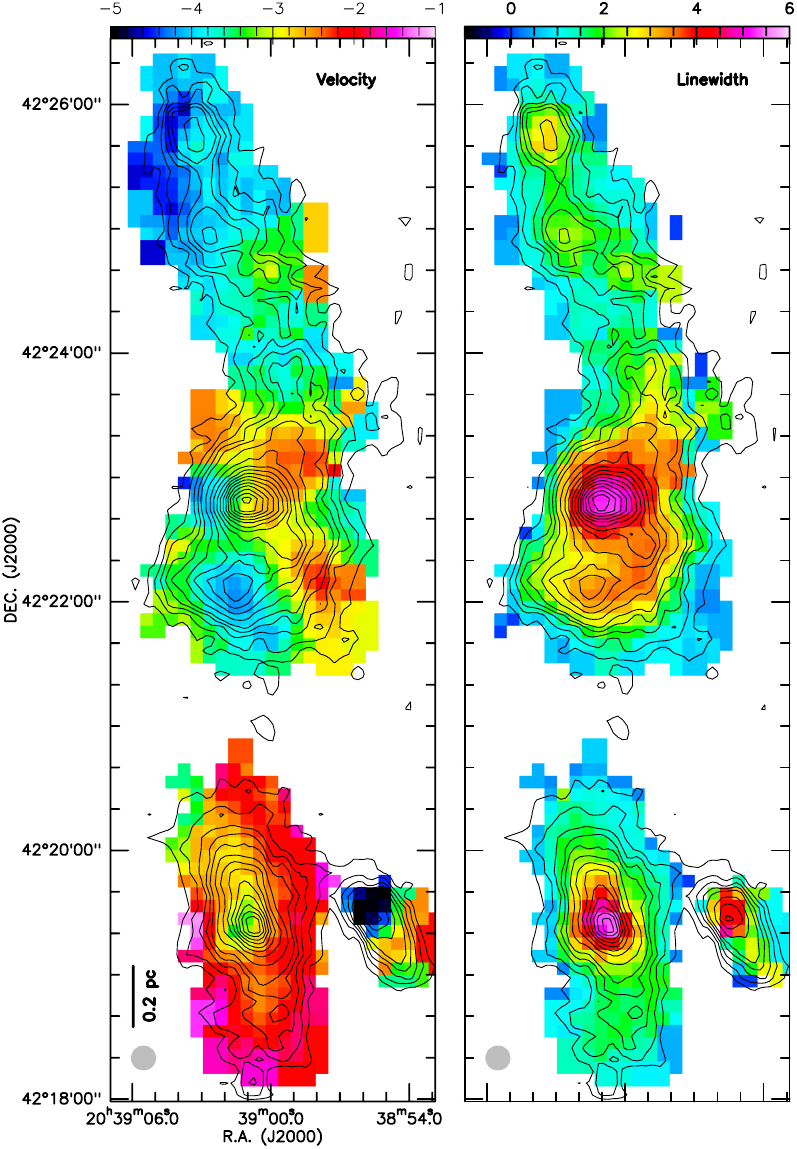}
\end{center}
\caption{The intensity-weighted velocity field (moment 1, \emph{left}) and linewidth (moment 2, \emph{right}) of H$_{2}$CO\,3$_{03}$--2$_{02}$.
The unit of each colour bar is km\,s$^{-1}$. Contours are H$_{2}$CO\,3$_{03}$--2$_{02}$ integrated intensities
(same as in Fig.\,\ref{fig:H2CO-intensity-maps}).}
\label{fig:H2CO-moment1&2}
\end{figure}

\begin{figure*}[ht!]
\centering
\includegraphics[width=0.98\textwidth]{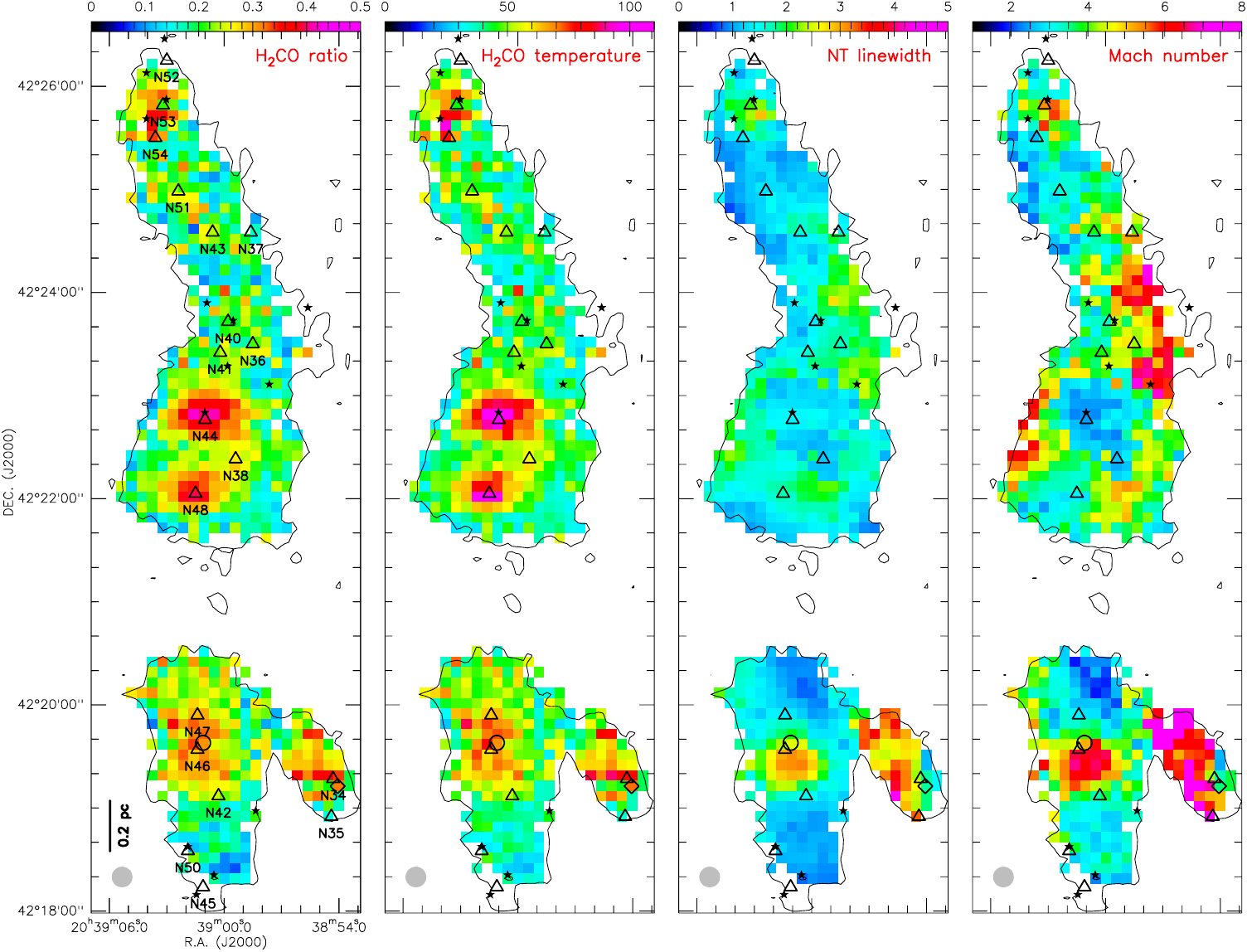}
\caption{Left two panels: The averaged velocity-integrated intensity ratio map of
H$_{2}$CO 0.5$\times$[(3$_{22}$--2$_{21}$\,+\,3$_{21}$--2$_{20})$/3$_{03}$--2$_{02}$]
in the DR21 filament (see Sect.\,\ref{sect:Results}). The kinetic temperatures derived from the
H$_{2}$CO line ratios (color bar in units of Kelvin).
Right two panels: Maps of non-thermal linewidth (color bar in units of km\,s$^{-1}$)
and Mach number. The black stars and triangles show
the locations of protostars \citep{Davis2007}
and the dense cores \citep{Motte2007}, respectively.
The black circle and diamond illustrate the locations of the explosion center as described by \cite{Zapata2013}
and the infrared-quiet massive dense core as delineated by \cite{Cao2019}, respectively.
Black contours show the integrated intensity of H$_{2}$CO\,3$_{03}$--2$_{02}$ at 4.8 K\,km\,s$^{-1}$
(see Fig.\,\ref{fig:H2CO-intensity-maps}).}
\label{fig:H2CO-ratio}
\end{figure*}

Pertaining to kinetic temperatures, it has been observed that subtle variations occur across different column densities
at a given H$_{2}$ number density of 10$^{5}$\,cm$^{-3}$ \citep{Tang2018a}. Specifically, we explored this relationship at column densities
of $N$(para-H$_{2}$CO)\,=\,1.0$\times$10$^{13}$, 6.6$\times$10$^{13}$, and 1.0$\times$10$^{14}$\,cm$^{-2}$, as depicted in Fig.\,\ref{fig:H2CO-radex}.
It indicates that there exists a slight temperature variation corresponding to different column densities.
The temperature differences derived from column densities of $N$(para-H$_{2}$CO)=1.0$\times$10$^{13}$ and 1.0$\times$10$^{14}$\,cm$^{-2}$,
compared to 6.6$\times$10$^{13}$\,cm$^{-2}$, are found to be smaller than the uncertainties in temperature determined from the errors
associated with observed H$_{2}$CO line ratios. Therefore we employed the established correlation between H$_{2}$CO line ratios and kinetic
temperature at a spatial density of 10$^{5}$\,cm$^{-3}$ and a column density of 6.6$\times$10$^{13}$\,cm$^{-2}$ in order
to convert the ratio maps into temperature maps in this work. These transformed temperature maps are illustrated in Fig.\,\ref{fig:H2CO-ratio}.
As previously stated, the measured H$_{2}$ number densities of massive dense cores within DR21 span from approximately $\sim$\,10$^{4}$ to 10$^{6}$\,cm$^{-3}$. Regarding kinetic temperatures, subtle variations have been noted across different number densities within the range of $\sim$\,10$^{4}$ to 10$^{6}$\,cm$^{-3}$ (see Fig.\,5 in \citealt{Tang2021}). It appears that at $n$(H$_2$)\,=\,10$^{5}$\,cm$^{-3}$, $T_{\rm kin}$ consistently exhibits lower values compared to those at 10$^{4}$ to 10$^{6}$\,cm$^{-3}$ by approximately $\lesssim$35\% for $T_{\rm kin}$\,$\lesssim$\,100\,K in Fig.\,\ref{fig:H2CO-ratio}.

\begin{figure*}[h]
\centering
\includegraphics[width=0.98\textwidth]{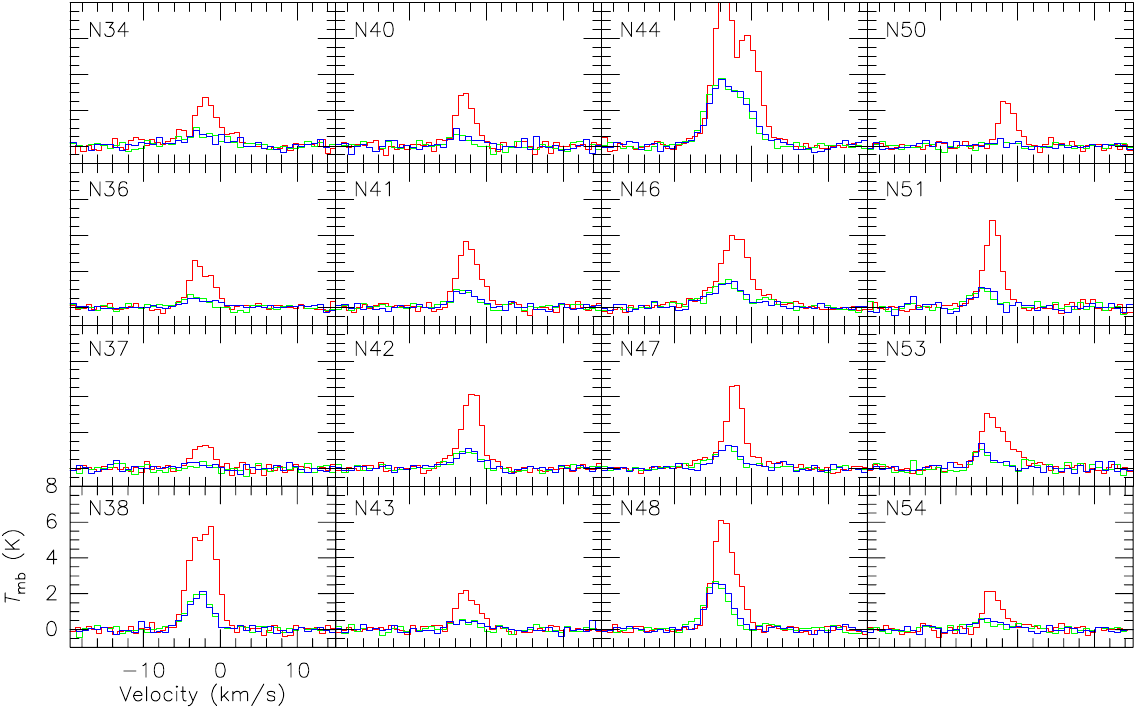}
\caption{Observed H$_{2}$CO spectra of the dense cores in the DR21 filament obtained with the IRAM\,30\,m.
Red, green, and blue lines show H$_{2}$CO\,3$_{03}$--2$_{02}$, 3$_{21}$--2$_{20}$, and 3$_{22}$--2$_{21}$, respectively.
The dense cores, identified through 1.2\,mm continuum emission as described by \cite{Motte2007}, are referenced for nomenclature,
positions, and physical parameters in Table\,\ref{table:Clumps-Parameters}.
}
\label{fig:H2CO-core}
\end{figure*}

The applicability of local thermodynamic equilibrium (LTE) as a reliable approximation for H$_{2}$CO level populations
has been established under certain conditions, namely when the corresponding transitions are optically thin and the targets are characterized by high densities
\citep{Mangum1993a}. By employing the methodology outlined in \cite{Tang2017b} (refer to their Eq.\,2) and utilizing the RADEX model,
we conducted a comprehensive analysis to investigate the relationship between kinetic temperature
and the combined H$_{2}$CO ratio within the LTE regime in Fig.\,\ref{fig:H2CO-radex}.
It indicates a satisfactory agreement between the temperatures obtained from the LTE model and those
derived from the RADEX model, under the condition of a spatial gas density of $n$(H$_2$)\,=\,10$^{5}$\,cm$^{-3}$,
as long as the kinetic temperature remains below 100\,K.

The kinetic temperature of the warm dense gas in the DR21 region, as indicated by the H$_{2}$CO line ratios, ranges from 24 to 114\,K,
with an average temperature of 48.3\,$\pm$\,0.5\,K at a density of $n$(H$_{2}$)\,=\,10$^{5}$\,cm$^{-3}$
(see also Table \ref{table:Parameters_compared}). As shown in Fig.\,\ref{fig:H2CO-ratio} and Table\,\ref{table:Temperatures},
it reveals the inhomogeneous
temperature distribution of the DR21 region across different sub-regions of the molecular complex.
The H$_{2}$CO line ratio analysis demonstrates that two dense clumps, DR21(OH) and DR21(OH)S,
located within the central parts of the DR21 filament, exhibit noticeably high temperatures
exceeding 100\,K. The kinetic temperatures of the dense clumps W75S\,FIR\,3, DR21, and DR21W1, albeit lower than in DR21(OH),
remain relatively high ranging between 77 to 84\,K. Conversely, the gas temperatures of the dense clumps W75S\,FIR\,1,
W75S\,FIR\,2, DR21(OH)N, DR21(OH)W, and DR21S span the range from 38 to 57\,K. Table\,\ref{table:Clumps-Parameters} provides
a summary of the detailed temperature results for all sixteen dense cores apparent in the DR21 filament.
In the immediate vicinity of the dense regions such as W75S\,FIR\,3, W75S\,FIR\,2, DR21(OH), DR21(OH)S, and DR21,
the kinetic temperatures range between 30 to 45\,K. The region located between DR21(OH) and W75S and the outskirts of the DR21
filament exhibit significantly lower temperatures below 40\,K.

To address the impact of H$_{2}$CO\,(3--2) line saturation on our temperature measurements, we employed the RADEX non-LTE model.
This model allowed us to calculate the opacities of H$_{2}$CO\,(3--2) lines, adopting an average temperature of $T_{\rm kin}$\,=\,48\,K,
an average measured linewidth of 3.8\,km\,s$^{-1}$, an H$_{2}$CO column density of $N$(para-H$_{2}$CO)\,=\,6.6$\times$10$^{13}$\,cm$^{-2}$,
and a gas volume density of $n$(H$_2$)\,=\,10$^{5}$\,cm$^{-3}$. The resulting optical depths for the
H$_{2}$CO\,3$_{03}$--2$_{02}$, 3$_{22}$--2$_{21}$, and 3$_{21}$--2$_{20}$ transitions were approximately 1.2, 0.28, and 0.28, respectively.
These values suggest that in the location of the dense cores, opacities in the H$_{2}$CO\,3$_{03}$--2$_{02}$ lines may exhibit slight saturation,
potentially leading to a slight overestimation of kinetic temperatures by $\sim$20$\%$.

\subsection{Thermal and Non-thermal Motions Obtained from H$_{2}$CO}
\label{sec:Th-non-th}
The thermal and non-thermal linewidths \citep{Pan2009,Dewangan2016}, represented by
$\sigma_{\rm T}$\,=\,$\sqrt{\frac{kT_{\rm kin}}{m_{\rm H_2CO}}}$ and
$\sigma_{\rm NT}$\,=\,$\sqrt{\frac{\Delta v^2}{8{\rm ln}2}-\sigma_{\rm T}^2}$\,$\thickapprox$\,$\Delta v/2.355$, respectively,
are determined based on the kinetic temperature derived from the H$_{2}$CO line ratios.
In this context, $k$ denotes the Boltzmann constant, $T_{\rm kin}$ corresponds to the kinetic temperature of the gas,
$m_{\rm H_2CO}$ represents the mass of a formaldehyde molecule, and $\Delta v$ signifies the measured FWHM linewidth
of H$_{2}$CO\,3$_{03}$--2$_{02}$. As mentioned in Sect.\,\ref{sect:Observations-and-data-reduction}, approximately 70 positions
in the DR21(OH) region exhibit two distinct velocity components at about $-$2 and $-$4\,\,km\,s$^{-1}$ (see also Fig.\,\ref{fig:H2CO-core}).
Additionally, determining
the individual gas temperatures for these positions in the two velocity components proved to be challenging. Hence, in this study,
we employ the approach of utilizing the averaged linewidth, which is weighted by the integrated intensity of H$_{2}$CO\,3$_{03}$--2$_{02}$.
By analyzing the data, the thermal linewidth ranges from 0.08 to 0.18\,km\,s$^{-1}$
with an average value of 0.11\,$\pm$\,0.01\,km\,s$^{-1}$, whereas the non-thermal linewidth spans from 0.74 to 4.40\,km\,s$^{-1}$
with an average value of 1.60\,$\pm$\,0.02\,km\,s$^{-1}$ (see also Table \ref{table:Parameters_compared}).
It is apparent that the non-thermal linewidth significantly exceeds the thermal linewidth, indicating that the dense
gas traced by H$_{2}$CO is primarily governed by non-thermal motions. These findings align with previous studies
conducted on star-forming regions in our Galaxy and the Large Magellanic Cloud using H$_{2}$CO\,($J$=3--2 and 4--3)
observations \citep{Tang2017a,Tang2018a,Tang2018b,Tang2021}.

The non-thermal linewidth distribution of H$_{2}$CO\,3$_{03}$--2$_{02}$ is depicted in Fig.\,\ref{fig:H2CO-ratio},
showing similarities with the H$_{2}$CO linewidth observed in the W75S and DR21 regions.
However, a significant disparity in the distribution of non-thermal linewidth and H$_{2}$CO linewidth
is observed in the DR21(OH) region, as illustrated in Fig.\,\ref{fig:H2CO-moment1&2}.
As mentioned above, the lower values for the non-thermal linewidth in the DR21(OH) region can be attributed
to the averaging of linewidths derived from the two distinct velocity components. In the western region of DR21(OH)N,
weak H$_{2}$CO emission is detected, as depicted in Fig.\,\ref{fig:H2CO-moment1&2}.
At many positions in this region, two velocity components are present, but one of them exhibits lower S/Ns,
making it challenging to identify it as a significant signal. Consequently, we employed a single Gaussian
velocity component to fit the H$_{2}$CO lines in this specific area, resulting in higher non-thermal linewidths
in the western region of DR21(OH)N.

\begin{table*}[h]
\scriptsize
\caption{Parameters of dense cores in DR21 filament.}
\centering
\begin{tabular}
{ccclcccccccccccccc}
\hline\hline
Dense core & Offset &Transition &$\int T_{\rm mb}$\,d$v$ &$V_{\rm LSR}$ &FWHM &$T_{\rm mb}$ & $\sigma_{\rm T}$ & $\sigma_{\rm NT}$ & $c_{\rm s}$  & $\mathcal{M}$ & $T_{\rm kin}$ & $T_{\rm turb}$ \\
       & " , " &  &K\,km\,s$^{-1}$ &km\,s$^{-1}$ &km\,s$^{-1}$ &K &km\,s$^{-1}$ & km\,s$^{-1}$ & km\,s$^{-1}$ & & K & K \\
\hline
N34 & --71.5, --211.5  &3$_{03}$--2$_{02}$                         &13.0 (0.8) &--2.4 (0.1) &5.4 (0.5) &2.3 & 0.15 (0.01) & 2.30 (0.20) & 0.54 (0.05) & 4.3 & 84 (15)  & 70 \\
    &                  &(3$_{21}$--2$_{20}$+3$_{22}$--2$_{21}$)/2  &4.6 (0.4)  &--2.4 (0.3) &6.9 (0.7) &0.6 &             &             &             &     &              &    \\
N36 & --29.5, 40.5     &3$_{03}$--2$_{02}$                         &10.2 (0.4) &--3.2 (0.1) &4.0 (0.2) &2.4 & 0.11 (0.01) & 1.68 (0.07) & 0.40 (0.02) & 4.2 & 46 (5)   & 43 \\
    &                  &(3$_{21}$--2$_{20}$+3$_{22}$--2$_{21}$)/2  &2.2 (0.2)  &--3.2 (0.2) &3.8 (0.5) &0.5 &             &             &             &     &              &    \\
N37 & --29.5, 106.5    &3$_{03}$--2$_{02}$                         &5.0 (0.4)  &--2.8 (0.1) &3.5 (0.3) &1.4 & 0.10 (0.01) & 1.47 (0.13) & 0.36 (0.04) & 4.0 & 38 (7)   & 36 \\
    &                  &(3$_{21}$--2$_{20}$+3$_{22}$--2$_{21}$)/2  &0.8 (0.2)  &--2.9 (0.4) &3.1 (0.8) &0.3 &             &             &             &     &              &    \\
N38 & --17.5, --25.5   &3$_{03}$--2$_{02}$                         &27.9 (2.5) &--2.9 (0.1) &2.7 (0.2) &4.8 & 0.13 (0.01) & 1.15 (0.07) & 0.44 (0.02) & 2.6 & 57 (6)   & 50 \\
    &                  &(3$_{21}$--2$_{20}$+3$_{22}$--2$_{21}$)/2  &7.5 (0.2)  &--2.9 (0.1) &3.0 (0.2) &1.2 &             &             &             &     &              &    \\
N40 & --11.5, 52.5     &3$_{03}$--2$_{02}$                         &9.9 (0.5)  &--3.5 (0.1) &3.1 (0.2) &3.0 & 0.11 (0.01) & 1.32 (0.08) & 0.39 (0.03) & 3.4 & 44 (7)   & 41 \\
    &                  &(3$_{21}$--2$_{20}$+3$_{22}$--2$_{21}$)/2  &2.0 (0.4)  &--3.5 (0.2) &2.8 (0.6) &0.7 &             &             &             &     &              &    \\
N41 & --11.5, 34.5     &3$_{03}$--2$_{02}$                         &14.5 (0.5) &--2.9 (0.1) &3.9 (0.2) &3.5 & 0.12 (0.01) & 1.64 (0.06) & 0.42 (0.01) & 3.9 & 51 (3)   & 47 \\
    &                  &(3$_{21}$--2$_{20}$+3$_{22}$--2$_{21}$)/2  &3.5 (0.2)  &--2.9 (0.1) &3.4 (0.2) &1.0 &             &             &             &     &              &    \\
N42 & --5.5, --223.5   &3$_{03}$--2$_{02}$                         &15.8 (0.4) &--2.4 (0.1) &3.5 (0.1) &4.3 & 0.12 (0.01) & 1.47 (0.04) & 0.43 (0.02) & 3.5 & 52 (4)   & 47 \\
    &                  &(3$_{21}$--2$_{20}$+3$_{22}$--2$_{21}$)/2  &3.9 (0.3)  &--2.5 (0.1) &3.6 (0.3) &1.0 &             &             &             &     &              &    \\
N43 & --5.5, 106.5     &3$_{03}$--2$_{02}$                         &8.6 (0.4)  &--3.1 (0.1) &3.8 (0.2) &2.1 & 0.12 (0.01) & 1.62 (0.08) & 0.42 (0.03) & 3.9 & 50 (8)   & 46 \\
    &                  &(3$_{21}$--2$_{20}$+3$_{22}$--2$_{21}$)/2  &2.0 (0.3)  &--2.7 (0.3) &3.9 (0.7) &0.5 &             &             &             &     &              &    \\
N44 & 0.5, --1.5       &3$_{03}$--2$_{02}$                         &54.7 (1.4) &--2.9 (0.1) &3.4 (0.1) &8.0 & 0.17 (0.01) & 1.43 (0.04) & 0.60 (0.04) & 2.4 & 103 (15) & 78 \\
    &                  &(3$_{21}$--2$_{20}$+3$_{22}$--2$_{21}$)/2  &21.8 (1.6) &--3.0 (0.6) &3.6 (0.6) &3.0 &             &             &             &     &              &    \\
N46 & 6.5, --193.5     &3$_{03}$--2$_{02}$                         &22.4 (0.6) &--2.8 (0.1) &5.5 (0.2) &3.9 & 0.15 (0.01) & 2.32 (0.08) & 0.52 (0.02) & 4.5 & 77 (7)   & 66 \\
    &                  &(3$_{21}$--2$_{20}$ +3$_{22}$--2$_{21}$)/2 &7.6 (0.4)  &--3.2 (0.1) &5.6 (0.4) &1.3 &             &             &             &     &              &    \\
N47 & 6.5, --175.5     &3$_{03}$--2$_{02}$                         &16.2 (0.5) &--2.8 (0.1) &3.3 (0.1) &4.7 & 0.13 (0.01) & 1.38 (0.06) & 0.47 (0.02) & 3.0 & 63 (5)   & 54 \\
    &                  &(3$_{21}$--2$_{20}$+3$_{22}$--2$_{21}$)/2  &4.7 (0.3)  &--3.2 (0.1) &3.7 (0.3) &1.2 &             &             &             &     &              &    \\
N48 & 6.5, --49.5      &3$_{03}$--2$_{02}$                         &26.1 (0.4) &--4.4 (0.1) &4.1 (0.1) &6.0 & 0.17 (0.01) & 1.73 (0.03) & 0.59 (0.02) & 2.9 & 100 (5)  & 77 \\
    &                  &(3$_{21}$--2$_{20}$+3$_{22}$--2$_{21}$)/2  &10.3 (0.2) &--4.4 (0.1) &3.6 (0.1) &2.7 &             &             &             &     &              &    \\
N50 & 6.5, --253.5     &3$_{03}$--2$_{02}$                         &7.8 (0.4)  &--1.8 (0.1) &2.9 (0.2) &2.6 & 0.09 (0.01) & 1.21 (0.07) & 0.32 (0.02) & 3.8 & 30 (4)   & 29 \\
    &                  &(3$_{21}$--2$_{20}$+3$_{22}$--2$_{21}$)/2  &0.9 (0.2)  &--2.2 (0.3) &3.1 (0.9) &0.3 &             &             &             &     &              &    \\
N51 & 18.5, 130.5      &3$_{03}$--2$_{02}$                         &15.2 (0.5) &--4.0 (0.1) &3.1 (0.1) &4.6 & 0.12 (0.01) & 1.32 (0.05) & 0.41 (0.02) & 3.2 & 48 (5)   & 44 \\
    &                  &(3$_{21}$--2$_{20}$+3$_{22}$--2$_{21}$)/2  &3.5 (0.4)  &--4.5 (0.2) &3.1 (0.3) &1.1 &             &             &             &     &              &    \\
N53 & 24.5, 178.5      &3$_{03}$--2$_{02}$                         &14.8 (0.6) &--3.8 (0.1) &5.0 (0.2) &2.8 & 0.11 (0.01) & 2.13 (0.10) & 0.40 (0.02) & 5.4 & 45 (5)   & 45 \\
    &                  &(3$_{21}$--2$_{20}$+3$_{22}$--2$_{21}$)/2  &3.2 (0.4)  &--4.6 (0.1) &2.6 (0.4) &1.1 &             &             &             &     &              &    \\
N54 & 30.5, 160.5      &3$_{03}$--2$_{02}$                         &7.2 (0.4)  &--4.0 (0.1) &3.1 (0.2) &2.2 & 0.15 (0.01) & 1.32 (0.10) & 0.52 (0.05) & 2.6 & 77 (16)  & 63 \\
    &                  &(3$_{21}$--2$_{20}$+3$_{22}$--2$_{21}$)/2  &2.4 (0.3)  &--3.9 (0.3) &4.8 (0.8) &0.5 &             &             &             &     &              &    \\
\hline
\end{tabular}
\label{table:Clumps-Parameters}
\tablefoot{Offsets relative to our reference position for the DR21 filament (see Sect.\,\ref{sect:Observations-and-data-reduction} and Fig.\,\ref{fig:H2CO-ratio}).
Velocity-integrated intensity, $\int T_{\rm mb}$\,d$v$, local standard of rest velocity, $V_{\rm LSR}$, full width
at half maximum line width (FWHM), and peak temperature ($T_{\rm mb}$) were obtained from Gaussian
fitting using CLASS as part of the GILDAS software package. For the thermal and non-thermal linewidths,
$\sigma_{\rm T}$ and $\sigma_{\rm NT}$, the sound velocity $c_{\rm s}$, the Mach number $\mathcal{M}$,
the kinetic temperature $T_{\rm kin}$, and the turbulent temperature $T_{\rm turb}$,
see Sects.\,\ref{sect:Results} and \ref{Sec:Turbulence}. Values in parentheses are uncertainties.}
\end{table*}

\begin{table*}[t]
\small
\caption{Thermal and non-thermal parameters derived from DR21, OMC-1, and N113.}
\centering
\begin{tabular}
{lccccccccccc}
\hline\hline
Parameter & \multicolumn{3}{c}{DR21 ($\sim$0.1\,pc}) & & \multicolumn{3}{c}{OMC-1$^a$ ($\sim$0.06\,pc}) & & \multicolumn{3}{c}{N113$^b$ ($\sim$0.4\,pc})\\
\cline{2-4} \cline{6-8} \cline{10-12}
& Range & Median & Mean & & Range & Median & Mean & & Range & Median & Mean\\
\hline
H$_{2}$CO line ratio              &0.08--0.42  &0.22 &0.22 $\pm$ 0.01 &&0.12--0.61 &0.21 &0.28 $\pm$ 0.01 &&0.10--0.38  &0.21 &0.22 $\pm$ 0.01\\
$T_{\rm kin}$ /K                  &23.9--114.5 &46.4 &48.3 $\pm$ 0.5  &&30--$>$200 &48.5 &62.0 $\pm$ 2.0  &&27.6--105.4 &48.5 &51.4 $\pm$ 0.4 \\
$\sigma_{\rm T}$ /\,km\,s$^{-1}$  &0.08--0.18  &0.11 &0.11 $\pm$ 0.01 &&0.09--0.27 &0.11 &0.12 $\pm$ 0.01 &&0.08--0.16  &0.11 &0.11 $\pm$ 0.01\\
$\sigma_{\rm NT}$ /\,km\,s$^{-1}$ &0.74--4.40  &1.47 &1.60 $\pm$ 0.02 &&0.34--2.78 &1.64 &0.98 $\pm$ 0.02 &&0.93--2.76  &1.64 &1.69 $\pm$ 0.01\\
$c_{\rm s}$ /\,km\,s$^{-1}$       &0.29--0.63  &0.40 &0.41 $\pm$ 0.01 &&0.30--0.97 &0.40 &0.44 $\pm$ 0.01 &&0.30--0.58  &0.40 &0.40 $\pm$ 0.01\\
$\mathcal{M}$                     &1.9--10.3   &3.7  &4.0  $\pm$ 0.1  &&0.7--4.3   &4.2  &2.3  $\pm$ 0.1  &&2.3--6.2    &4.2  &4.2  $\pm$ 0.1 \\
\hline 
\end{tabular}
\label{table:Parameters_compared}
\tablefoot{$^{(a)}$ Parameters of OMC-1 taken from \cite{Tang2018a}. $^{(b)}$ Parameters of N113 taken from \cite{Tang2021}.
The values enclosed in parentheses denote the linear scale of each respective source. The errors shown in Columns 4 and 7 are the standard
deviations of the mean.}
\end{table*}

Distributions of the Mach number
(given as $\mathcal{M}$\,=\,$\sigma_{\rm NT}/a_{\rm s}$, $a_{\rm s}$\,=\,$\sqrt{\frac{kT_{\rm kin}}{\mu m_{\rm H}}}$,
where $\mu$\,=\,2.37 is the mean molecular weight for molecular clouds and $m_{\rm H}$ is the mass of a hydrogen atom)
of the DR21 filament is shown in Fig.\,\ref{fig:H2CO-ratio}. The Mach number ranges from 1.9 to 10.3 with an average of
4.0\,$\pm$\,0.1 (see also Table\,\ref{table:Parameters_compared}). This suggests that supersonic non-thermal motions
(e.g., turbulence, outflows, shocks, and/or magnetic fields) are dominant in the dense gas traced by H$_{2}$CO.
The DR21W1 region, linked to the western DR21 flow, exhibits particularly high Mach numbers ($\mathcal{M}$\,$\sim$\,10)
as evidenced by previous studies (e.g., \citealt{Jaffe1989,Lane1990,Davis2007,Zapata2013}). This association indicates
a strong influence of shocks on the dense gas traced by H$_{2}$CO. Within the dense core of the DR21 region,
the Mach numbers are approximately $\mathcal{M}$\,$\sim$\,7, while in the W75S\,FIR\,3 region,
they are around $\mathcal{M}$\,$\sim$\,6, slightly lower than those observed in the DR21W1 region.
This observation of high Mach numbers suggests that local star formation activities
(e.g., outflows, shocks; \citealt{Davis2007,Motte2007,Schneider2010})
in the DR21 and W75S\,FIR\,3 regions may impact the dense gas. In the DR21(OH) region, the Mach numbers are lower (2--3),
whereas in the western region of DR21(OH)N,
they are higher (6--7). These variations are attributed to the two velocity components mentioned in the above paragraph.
Moving towards the outskirts and extended regions of the DR21 filament, the typical Mach number ranges from 2 to 4.

\section{Discussion}
\label{sect:discussion}
\subsection{Comparison of Temperatures Derived from H$_{2}$CO, NH$_3$, and CO}
\label{Sec:Tk-H2CO-others}
Ammonia (NH$_{3}$) is widely recognized as a standard temperature tracer for dense molecular gas (e.g., \citealt{Ho1983}).
Its unique properties make it an invaluable tool for studying the thermal characteristics of various molecular environments
(e.g., \citealt{Henkel1987,Henkel2008,Molinari1996,Jijina1999,Dunham2011,Wang2012,Wienen2012,Mangum2013b,Lu2014,Lu2017,Ott2014,Gong2015a,Gong2015b,Krieger2017,Sokolov2018,Wang2024,Zhang2024}).
In order to investigate the temperature variations across various layers of the molecular cloud, we conducted a comparative
analysis between our results of gas temperature derived from H$_{2}$CO\,(3--2)
(with $n_{\rm crit}$(H$_2$CO\,3$_{03}$--2$_{02}$)\,$\sim$\,few\,10$^{5}$\,cm$^{-3}$, \citealt{Shirley2015})
and the previous measurements using NH$_3$\,$(J,K)$\,=\,(1,1) and (2,2)
(with $n_{\rm crit}$(NH$_{3}$)\,$\sim$\,10$^{3}$\,cm$^{-3}$, \citealt{Shirley2015}), specifically within the DR21 filament.
As mentioned in Sect.\,\ref{sect:DR21-filament}, the temperature distribution of gas within the DR21 filament has been charted
using NH$_3$\,(1,1) and (2,2) with the GBT (beam size $\sim$32$''$; \citealt{Keown2019}). Analysis of the NH$_3$\,(2,2)/(1,1)
line ratios has shown that the kinetic temperature of the dense gas typically ranges between 20 and 30\,K. However, there are variations
in temperature across different regions (see Table\,\ref{table:Temperatures}). Specifically, the W75S region
has a typical temperature of about 20\,K, while the DR21(OH)
region has temperatures ranging from 22 to 64\,K, and the DR21W1 region has temperatures around 35--60\,K. High-resolution imaging
of the NH$_3$\,(1,1) and (2,2) lines within the DR21(OH) region was presented by \cite{Mangum1992}, using the VLA with a beam size
of $\sim$4$''$ ($\sim$0.03\,pc at 1.5\,kpc distance). The gas kinetic temperatures of dense ammonia cores derived from line ratios
of NH$_3$\,(2,2)/(1,1) range from 20 to >80 K, with a typical value of $\sim$30\,K.
On a larger scale of $\sim$0.23\,pc, the temperature distribution derived from our H$_{2}$CO\,(3--2)
data exhibits similarities to that obtained from NH$_3$\,(1,1) and (2,2). Notably, our analysis of H$_{2}$CO\,(3--2) consistently
reveals higher gas temperatures compared to NH$_3$\,(1,1) and (2,2), both within dense clumps and across extended regions of the
DR21 filament (see Sect.\,\ref{sec:Tk-H2CO} and Table\,\ref{table:Clumps-Parameters}). These findings align with previous observations
in star-forming regions of the Milky Way, the Large Magellanic Cloud, and our Galactic center clouds
\citep{Tang2017a,Tang2017b,Tang2018a,Tang2018b,Tang2021,Ginsburg2016}.

Observations conducted using the KOSMA 3\,m telescope have provided insights into the gas properties within the DR21 filament region
through the analysis of low- to high-$J$ CO lines on a spatial scale of $\sim$0.4\,pc \citep{Jakob2007}.
These observations reveal the presence of two distinct gas components (see Table\,\ref{table:Temperatures}).
The cold component exhibits a kinetic temperature ranging
from 30 to 40\,K, while the warm component displays temperatures ranging from 80 to 150\,K in regions such as W75S\,FIR\,1, DR21(OH),
DR21, and DR21W1, at densities of $n$(H$_{2}$)\,$\sim$\,10$^{4}$--10$^{6}$\,cm$^{-3}$. Furthermore, in these regions,
the gas temperatures derived from H$_{2}$CO line ratios range from 50 to 114\,K on a spatial scale of $\sim$0.1\,pc. Notably,
these values align closely with the temperatures obtained from the warm gas component identified through the analysis of CO lines.

The observed disparities likely arise from temperature gradients, as H$_{2}$CO and NH$_3$ may probe distinct regions within
star-forming regions. The H$_{2}$CO\,(3--2) lines emerge as an interesting tool for detecting the presence of warm and dense
molecular gas within star-forming regions. These transitions are highly sensitive to environments where the gas temperature
is elevated, making it an ideal tracer of energetic conditions associated with active star formation. In contrast,
the NH$_3$\,(1,1) and (2,2) transitions are well-suited for probing the cooler and less dense molecular gas prevalent
in these regions. It would be intriguing to conduct mapping of the region using higher excited NH$_3$ transitions
(e.g., \citealt{Henkel1987,Henkel2008,Mangum2013b,Lu2017,Gong2015a,Gong2015b,Krieger2017}).
This approach would also serve as a valuable tool to verify the authenticity of the high temperatures, ensuring that they are
not a result of slightly saturated H$_{2}$CO\,3$_{03}$--2$_{02}$ lines.

\begin{table*}[t]
\caption{Comparison of temperatures and Mach number of different regions.}
\centering
\begin{tabular}
{lccccc}
\hline\hline 

Region &H$_2$CO &NH$_3$ &CO &FIR &Mach number\\
&K&K&K&K&\\
\hline 
W75S      &24--95   &15--64    &41(82/37)   &15--25    &2.1--7.2\\
DR21(OH)  &26--114  &22--64    &31(99/29)   &15--31    &2.2--7.0\\
DR21      &24--87   &20--58    &36(148/32)   &15-->22   &1.9--6.7 \\
DR21W1    &33--88   &35--60    &54(118/52)  &18--22    &2.5--10.3 \\
\hline 
\end{tabular}
\tablefoot{Column 2: kinetic temperatures derived from H$_2$CO line ratios (see Sect.\,\ref{sec:Tk-H2CO}).
Column 3: kinetic temperatures derived from the NH$_3$\,(2,2)/(1,1) line ratios taken from \cite{Keown2019}.
Column 4: average, warm, and cold components temperatures of low-- to high--$J$ CO taken from \cite{Jakob2007}.
Columns 5: far infrared (FIR) dust temperatures taken from \cite{Hennemann2012} and \cite{Cao2019}.
Columns 6: Mach number derived from H$_{2}$CO (see Sect.\,\ref{sec:Th-non-th}).}
\label{table:Temperatures}
\end{table*}

\subsection{Comparison of Temperatures Obtained from Gas and Dust}
\label{Sec:Tk-H2CO-dust}
Gas temperature refers to the kinetic temperature of the gas particles, while dust temperature reflects the thermal
equilibrium of the dust grains. The investigation of gas and dust temperatures in molecular clouds is of
paramount importance in understanding the physical conditions within these regions.
It is commonly expected that gas and dust are thermally coupled in dense regions
($n$(H$_2$)\,$\gtrsim$\,10$^{4}$\,cm$^{-3}$) due to the increased frequency of interactions between these two
components \citep{Goldsmith2001}. Previous observations with NH$_3$\,(1,1) and (2,2) lines have consistently
demonstrated that the temperatures derived from gas and dust (derived at far-infrared wavelengths)
measurements are typically in concordance within the active dense clumps found in Galactic disk clouds
(e.g., \citealt{Dunham2010,Giannetti2013,Battersby2014,Merello2019,Tursun2020,Zhang2024}). However, it should be noted
that gas temperatures obtained from H$_{2}$CO\,(3--2 and 4--3) lines do not exhibit agreement with
FIR-derived dust measurements in both our Galactic star-forming regions \citep{Tang2017a,Tang2018a,Tang2018b}
and Galactic center clouds \citep{Ao2013,Ginsburg2016,Immer2016}.

As detailed in Sect.\,\ref{sect:DR21-filament}, the dust temperature distribution within the DR21 filament
has been established through analyzing {\it Herschel} continuum data observed at 70--500\,$\mu$m (beam size $\sim$40$''$),
as reported by both \cite{Hennemann2012} and \cite{Cao2019}. The analysis reveals that the dense clumps
within our observed region of the DR21 filament exhibit a range of dust temperatures spanning from 15 to 35\,K
(see Table\,\ref{table:Temperatures}).
The typical value, approximately 20\,K, is found to be similar to the gas temperature derived from the
NH$_3$\,(2,2)/(1,1) line ratios \citep{Mangum1992,Keown2019}, albeit lower than the results obtained from
the H$_{2}$CO line ratios (see Sect.\,\ref{sec:Tk-H2CO} and Table\,\ref{table:Clumps-Parameters}).
These findings align with previous observational results in star-forming regions within our
galaxy \citep{Tang2017a,Tang2018a,Tang2018b}.

In order to facilitate a comparison between the temperatures derived from gas and dust on a larger scale, we have
applied a smoothing technique to our H$_{2}$CO data, resulting in a beam size of $\sim$40$''$. The peak temperatures
obtained from this analysis are approximately 80, 95, 80, and 75\,K in the W75S\,FIR\,3, DR21(OH), DR21(OH)S, and DR21 regions,
respectively. It is worth noting that these values remain higher than the temperatures measured using NH$_3$ and dust emission
at FIR wavelengths. The structure of dust temperature closely resembles that of the gas temperature obtained
through NH$_3$ and H$_{2}$CO observations on a scale of $\sim$0.3\,pc in the DR21 filament. It is important
to note that the emission of dust at FIR wavelengths primarily originates from colder dust components which
may not be directly associated with the process of star formation (e.g., \citealt{Mangum2013b,Tang2018a}).
In comparison to temperature measurements utilizing NH$_3$\,(1,1) and (2,2) and the FIR continuum, the H$_{2}$CO lines
trace distinctly higher temperatures. This observation confirms that H$_{2}$CO may possess sensitivity
towards the dense gas that is intimately linked to active star formation.

\subsection{Gas Heating}
Gas heating in star-forming regions is a multifaceted process that holds paramount significance in the formation
and evolution of stars. Acquiring a comprehension of the pivotal role played by gas heating in
these regions, measurements of temperature are mandatory to elucidate the underlying mechanisms that drive star formation and shape the
evolutionary trajectory of galaxies.

\subsubsection{Radiative Heating}
\label{Sec:Radiation-heating}
Radiative heating plays a pivotal role in the intricate process of star formation, particularly during
the initial phases when protostars coalesce within dense molecular clouds. During this period,
protostars accrete material and emit substantial amounts of radiation, which interact with and
profoundly influence the surrounding ISM. Radiation transfers energy to the ISM
through processes such as photoionization, photodissociation, and the generation of photoelectrons.
These energy transfer mechanisms result in the heating of the ISM and serve as driving forces behind
the subsequent stages of star formation. Previous studies have conducted observations of several molecules,
such as NH$_3$, H$_{2}$CO, H$_{2}$CS, CH$_3$CN, CH$_3$CCH, and CH$_3$OH, in regions where massive stars are forming
(e.g., \citealt{Lu2014,Giannetti2017,Tang2018a,Tang2018b,Tang2021,Gieser2019,Gieser2021,Gieser2022,Gieser2023,Lin2022,Wang2023}).
These observations suggest that the dense gas surrounding embedded infrared sources experiences
internal radiative heating on scales of $\lesssim$0.1--1.0\,pc. As stated in Sects.\,\ref{sec:H2CO-Ratio}
and \ref{sec:Tk-H2CO}, high gas temperatures obtained from H$_{2}$CO line ratios associate with dense clumps
especially in the W75S\,FIR\,3, DR21(OH), DR21(OH)S, and DR21 regions (see Fig.\,\ref{fig:H2CO-ratio}).
This indicates that the warm dense gas traced by H$_{2}$CO may be influenced by radiation from internally
embedded infrared sources. Furthermore, investigations of CO, [CI], [CII], and [OI] emissions
in the DR21 region have revealed that the gas experiences significant
radiative heating \citep{Lane1990,Jakob2007,Ossenkopf2010}.

\begin{figure*}[t]
\vspace*{0.2mm}
\begin{center}
\includegraphics[width=1.0\textwidth]{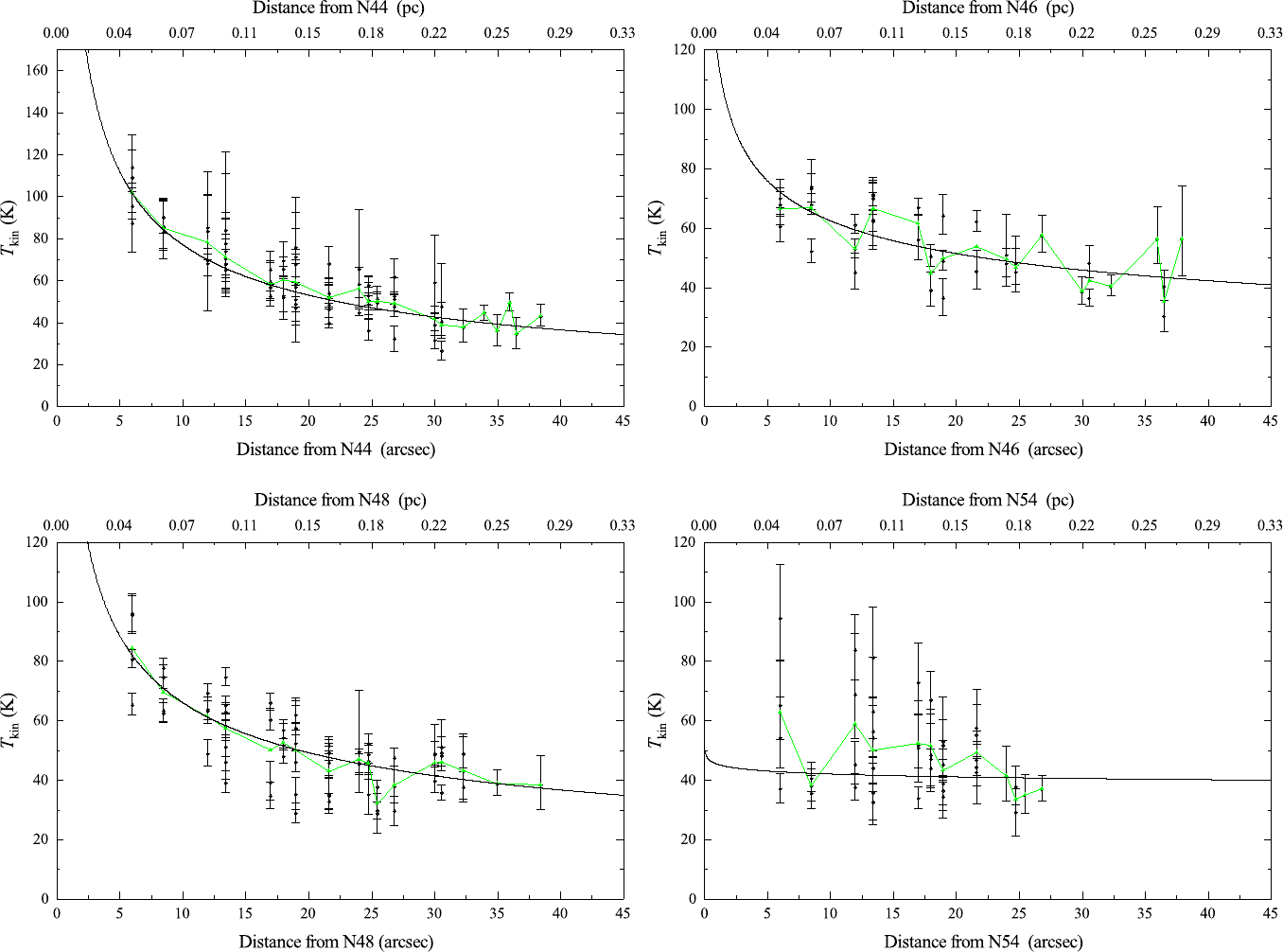}
\end{center}
\caption{Gas kinetic temperatures were determined by analyzing the H$_{2}$CO\,(3--2) line ratios from the central regions
of the dense cores N44, N46, N48, and N54 (see Table\,\ref{table:Clumps-Parameters}), extending towards their respective edges.
The format of the regression fits is $T_{\rm kin}(\rm cores) = a\times(\frac{R}{\rm arcsec})^{b}~{\rm K}$ while the fitting
method is the Levenberg-Marquardt method without the error of temperature, and the Reduced Chi-Squared values for N44, N46, N48,
and N54 are 1.7, 3.3, 5.7, and 1.9, respectively.
The average kinetic temperature at each projected distance is represented by green triangles. The green line represents the
connection of the each average kinetic temperature, while the fitted results for H$_{2}$CO are depicted by black lines.}
\label{fig:Tk-distance}
\end{figure*}

To investigate the influence of radiation heating on the extended dense gas surrounding massive clumps
within the DR21 filament on a scale of $\sim$0.1--0.3\,pc, we conducted a temperature profile analysis
of four dense cores situated in the W75S\,FIR\,3, DR21(OH), DR21(OH)S, and DR21 regions
(see Fig.\,\ref{fig:Tk-distance}). These cores, namely N44, N46, N48, and N54, chosen from the
respective regions, exhibit distinctly higher kinetic temperatures derived from H$_{2}$CO line ratios
(see Table\,\ref{table:Clumps-Parameters}) than those of the surrounding gas.
We note that Fig.\,\ref{fig:H2CO-ratio} shows the spatial distribution of the dense core N38 to
the west of N44 and N48 ($\sim$32$''$, $\sim$0.24\,pc), the dense core N42 to the southwest of N46
($\sim$32$''$, $\sim$0.24\,pc),  the dense core N47 to the north of N46 ($\sim$18$''$, $\sim$0.14\,pc),
and the dense core N53 to the north of N54 ($\sim$19$''$, $\sim$0.14\,pc). Due to the limited
resolution of our observations (beam size $\sim$12$''$), we cannot rule out that nearby sources may
have influenced the temperature profiles of N44, N46, N48, and N54. To reduce this influence from
nearby dense cores, we masked data at the midpoint between selected dense cores and other sources.
For the N44 dataset, southern data was masked using a distance threshold exceeding 19$''$ along
the declination axis. In the case of the N46, both northern and southern data were masked at
distances greater than 9$''$ and 15$''$, respectively. For N48, northern data was masked at a
distance exceeding 19$''$. Similarly, for N54, northern data was masked using a distance exceeding 10$''$.
Additionally, for N44 and N48, data corresponding to the dense core N38 (with a radius of $\sim$16$''$) was also masked.

In the dense cores N44, N46, and N48, there are significant temperature gradients observed,
as depicted in Fig.\,\ref{fig:Tk-distance}. However, it is worth mentioning
that a relatively shallow temperature gradient is detected in dense core N54, which could potentially
be attributed to the fact that the dense core is likely in an earlier evolutionary stage (e.g., \citealt{Motte2007,Cao2019}),
with the embedded protostars or clusters not yet generating sufficient heat to significantly warm the
surrounding molecular envelope.
As noted above, the dense core N53 is positioned to the north of N54, at a distance of
approximately 19$''$ ($\sim$0.14\,pc). Previous observations have detected the presence of CH$_3$OH
and H$_2$O masers in N53 (e.g., \citealt{Motte2007}), suggesting the occurrence of significant shocks in this area.
Additionally, two protostars were identified in the N53/N54 region by \cite{Davis2007} (see Fig.\,\ref{fig:H2CO-ratio}).
The complex temperature structure observed in the N53/N54 region may be attributed to the impact of outflows and shocks
originating from the embedded protostars or clusters within N53 and N54, as well as the radiation emitted by nearby protostars.
The temperature profile surrounding these dense clumps can be described by the Stefan-Boltzmann blackbody radiation law,
expressed as $T_{\rm kin}=0.86\times(\frac{L}{{\rm L}_{\odot}})^{1/4}(\frac{R}{\rm pc})^{-1/2}~{\rm K}$,
where the luminosity $L$ is given in terms of solar luminosity (${\rm L}_{\odot}$) and the distance $R$
is measured in parsecs. By adjusting the emissivity of dust grains to be smaller than the characteristic
black-body temperature wavelength, the radiation law is modified to
$T_{\rm kin}=2.7\times(\frac{L}{{\rm L}_{\odot}})^{1/5}(\frac{R}{\rm pc})^{-2/5}~{\rm K}$ (\citealp{Wiseman1998,Tang2018a}).
The fitted results of the four dense cores are
\begin{eqnarray}
\label{equation:N44}
T_{\rm kin} ({\rm N44}) = (268\pm30) \times \bigg(\frac{R}{\rm arcsec}\bigg)^{-0.54\pm0.04}~{\rm K},
\end{eqnarray}
\begin{eqnarray}
\label{equation:N46}
T_{\rm kin} ({\rm N46}) = (120\pm16) \times \bigg(\frac{R}{\rm arcsec}\bigg)^{-0.28\pm0.05}~{\rm K},
\end{eqnarray}
\begin{eqnarray}
\label{equation:N48}
T_{\rm kin} ({\rm N48}) = (176\pm21) \times \bigg(\frac{R}{\rm arcsec}\bigg)^{-0.43\pm0.04}~{\rm K},
\end{eqnarray}
\begin{eqnarray}
\label{equation:N54}
T_{\rm kin} ({\rm N54}) = (45\pm11) \times \bigg(\frac{R}{\rm arcsec}\bigg)^{-0.03\pm0.09}~{\rm K},
\end{eqnarray}
with power-law indices of $-$0.54, $-$0.28, $-$0.43, and $-$0.03 for N44, N46, N48, and N54, respectively.
With the exception of the dense core N54, the temperature power-law indices obtained from fitting the H$_{2}$CO
line ratios in our study exhibit consistency with previous observational results with a mean value of about $-$0.4.
These results align with previous studies that have utilized various molecular tracers such as H$_{2}$CO,
H$_{2}$CS, CH$_3$CN, and CH$_3$CCH in the investigation of massive star-forming dense cores
(e.g., \citealt{Gieser2019,Gieser2021,Gieser2022,Gieser2023,Lin2022}). This indicates that the dense gas heating
of the massive cores on a scale of $\sim$0.1--0.3\,pc in the DR21(OH), DR21(OH)S, and DR21 regions is
significantly influenced by internal radiation emitted by associated protostar(s) and/or stellar clusters.
The resolution of our observation, $\sim$0.1\,pc, does not allow for a thorough measurement of the detailed temperature
structures within these dense cores. Consequently, a future high-resolution observation is necessary to address this limitation.

\subsubsection{Explosive Heating}
\label{Sec:Explosive-Heating}
Explosive heating in molecular clouds refers to the process by which a large amount of energy is rapidly released, leading to
a sudden increase in temperature within these clouds. The increased temperature can trigger chemical reactions,
influence the stability of the cloud, and even initiate the formation of new stars. It may play a crucial role in
the dynamics and evolution of molecular clouds. Understanding the processes and effects of explosive heating
is essential for studying the complex interplay between gas dynamics, star formation, and the overall
evolution of molecular clouds. Previous observations of H$_{2}$CO and H$_{2}$CS in the Orion\,KL region have
provided evidence that the dense gas in this area experiences heating due to shocks induced by a mysterious
explosive event that took place several hundred years ago (e.g., \citealt{Tang2018a,Li2020}). This event had a
significant impact on the energetics of the Orion\,KL region. \cite{Zapata2013} proposed that the enigmatic
northeast–southwest outflow observed in the DR21 region was likely the result of an explosive event that transpired $\sim$10,000
years ago. This event bears resemblance to the disintegration of a massive stellar system, akin to the
occurrence witnessed in Orion\,KL 500 years ago, albeit with approximately 20 times greater energy output.
The accurate location of this explosive event in the DR21 region was identified at
$\alpha_{2000}$\,=\,20$^h$39$^m$01\hbox{$\,.\!\!^s$}1 and $\delta_{2000}$\,=\,42$\degr$19$'$37\hbox{$\,.\!\!^{\prime\prime}$}9
by \cite{Zapata2013}. At the location of the DR21 outflow origin, there is no presence of a young star in the radio, submillimeter,
or infrared wavelengths. This absence suggests that the potential "source" of the outflow may no longer be situated there,
similar to the explosive outflow observed in Orion\,KL \citep{Zapata2009,Zapata2013}.

The explosive event is observed to occur at the center of an expanding cometary H\,{\scriptsize II} region in DR21
(see Fig.\,3 in \citealt{Zapata2013}). As previously mentioned, the heating of dense gas in the DR21 region is greatly
influenced by the internal radiation emitted by the associated protostar(s) and/or stellar clusters.
In the star formation region of DR21, differentiating between radiation and explosive heating can be challenging
but can be done by considering the detailed spatial distribution of $T_{\rm kin}$. Radiation heating is usually
more evenly distributed throughout a star-forming region, as it originates from the young stars themselves.
Explosive heating, however, may exhibit localized regions of enhanced heating, corresponding to the locations of
energetic events (e.g., \citealt{Tang2018a,Li2020}). Previous observations suggest that H$_{2}$CO could serve as
a promising indicator for detecting outflows or shocks (e.g.,\,\citealt{Tang2017b,Lu2021,Izumi2023}).
As stated in Sect.\,\ref{sec:Th-non-th}, the DR21 region exhibits substantial non-thermal
linewidths ($\sigma_{\rm NT}$\,$\sim$\,2.3\,km\,s$^{-1}$) and Mach numbers
($\mathcal{M}$\,$\sim$\,4.5), as depicted in Fig.\,\ref{fig:H2CO-ratio}. Their correlation suggests
that the dense gas, which is traced by H$_{2}$CO, may be influenced by the shocks
resulting from explosive flows and/or expanding H\,{\scriptsize II} regions.
To investigate the impact of the explosive event on the heating of
dense gas in the DR21 region, we conducted an analysis of the temperature structure within the dense core N46
(see Figs.\,\ref{fig:H2CO-ratio} and \ref{fig:Tk-distance}) which is located at the putative position of the
supposed explosive event \citep{Zapata2013}. The analysis of H$_{2}$CO data reveals
that the gas temperature within the dense core N46 of the DR21 region is measured to be 77\,K.
This temperature is lower than that observed in the dense cores N44 and N48 of the DR21(OH)
and DR21(OH)S regions. Furthermore, the temperature gradient observed in the DR21 region
is flatter when compared to the temperature gradients observed in the DR21(OH) and DR21(OH)S regions
(see Fig.\,\ref{fig:Tk-distance}). As previously noted, the dense gas may exhibit localized regions
of increased heat when subjected to heating by an explosive event. However, a relatively uniform temperature
gradient in the DR21 region is inconsistent with this pattern of explosive heating. As delineated in the preceding section,
the temperature profile derived from the dense core N46 aligns with the results anticipated from radiation heating.
The heating of the dense gas in the DR21 region may be partly attributed to the shocks generated by an explosive event
that occurred a long time ago. However, our temperature measurements with H$_{2}$CO do not provide direct evidence
to support the hypothesis that the dense gas is heated by shocks resulting from a past
explosive event on a scale of $\sim$0.1\,pc in the DR21 region.

\cite{Zapata2013} proposed that the DR21 northeast–southwest outflow was likely generated by a powerful explosion
occurring $\sim$10,000 years ago. However, the genesis of the DR21 outflow remains a subject of ongoing investigation
(e.g., \citealt{Skretas2023}). Previous observations suggest that the DR21W1 region is correlated with the western DR21 flow
(e.g., \citealt{Jaffe1989,Lane1990,Wilson1990,Garden1991a,Garden1991b,Garden1992,Davis2007}).
It appears that a high temperature ($\sim$84\,K) associates with the dense core N34 in the DR21W1 region
(see Fig.\,\ref{fig:H2CO-ratio}). A flat temperature structure, analogous to that of the DR21 region,
was detected in the DR21W1 region. The highest non-thermal linewidths
($\sigma_{\rm NT}$\,$\sim$\,4.4\,km\,s$^{-1}$) and Mach numbers ($\mathcal{M}$\,$\sim$\,10) of the entire
DR21 filament are found in the northeastern and southern region of DR21W1 (see Fig.\,\ref{fig:H2CO-ratio}).
These confirm that the dense gas in the DR21W1 region is strongly influenced by shocks which may be caused by explosive flows.
In the DR21W1 region, \cite{Cao2019} discovered an infrared-quiet massive dense core with a
luminosity of $\sim$350\,${\rm L}_{\odot}$. The absence of strong 8\,$\mu$m emission in this
region (refer to Fig.\,1 in \citealt{Kumar2007}) suggests that the dense core is in an
early evolutionary stage. In the study conducted, the gas temperature in the DR21W1 region,
which is influenced by young stellar objects (YSOs), was determined using the modified Stefan-Boltzmann
blackbody radiation law (refer to Sect.\,\ref{Sec:Radiation-heating}). It was assumed that the embedded
YSOs (or YSO candidates) serve as the primary sources of energy in the DR21W1 region, with a luminosity
of $\sim$350\,${\rm L}_{\odot}$ as reported by \cite{Cao2019}. The derived gas temperature from the modified
Stefan-Boltzmann blackbody radiation law was found to be $\sim$22\,K at radii of 0.1\,pc, which is significantly
lower than the typical temperature of 70--80\,K obtained from the measurements of H$_{2}$CO line ratios.
This indicates that the radiation heating from internal embedded YSOs (or YSO candidates) only affects
the local region. The high temperatures of dense gas in the DR21W1 region should be dominated by the shocks
from the western DR21 flow. Our measurements of H$_{2}$CO towards the DR21W1 region provide compelling
evidence that the dense gas in this area is heated by shocks which may arise from explosive flows.

\subsubsection{Turbulent Heating}
\label{Sec:Turbulence}
Turbulence is a prevalent phenomenon observed in various phases of the interstellar medium
\citep{Maclow2004,Elmegreen2004,Scalo2004,Sokolov2018,Wang2023,Wang2024}.
Recent observations conducted on Galactic Central Molecular Zone clouds using H$_{2}$CO\,(3--2 and 4--3)
indicate that the warm dense gas is primarily heated by turbulence occurring on a scale of $\sim$1\,pc
\citep{Ao2013,Ginsburg2016,Immer2016}. In Galactic massive star-forming regions, it appears that molecular gas heated by turbulence is
commonly encountered \citep{Tang2018a,Tang2018b,Tang2021}. Notably, multiple observations utilizing molecular tracers, including H$_{2}$CO,
NH$_{3}$, and CH$_{3}$CCH, in star formation regions, indicate a significant correlation between linewidth and gas kinetic temperature \citep{Wouterloot1988,Molinari1996,Jijina1999,Wu2006,Urquhart2011,Urquhart2015,Wienen2012,Lu2014,Giannetti2017,Tang2017a,Tang2018a,Tang2018b,Tang2021}.
As a result, higher temperatures inferred from various tracers align with elevated turbulence levels.

\begin{figure}[ht!]
    \begin{center}
    \includegraphics[scale=0.5,angle=0]{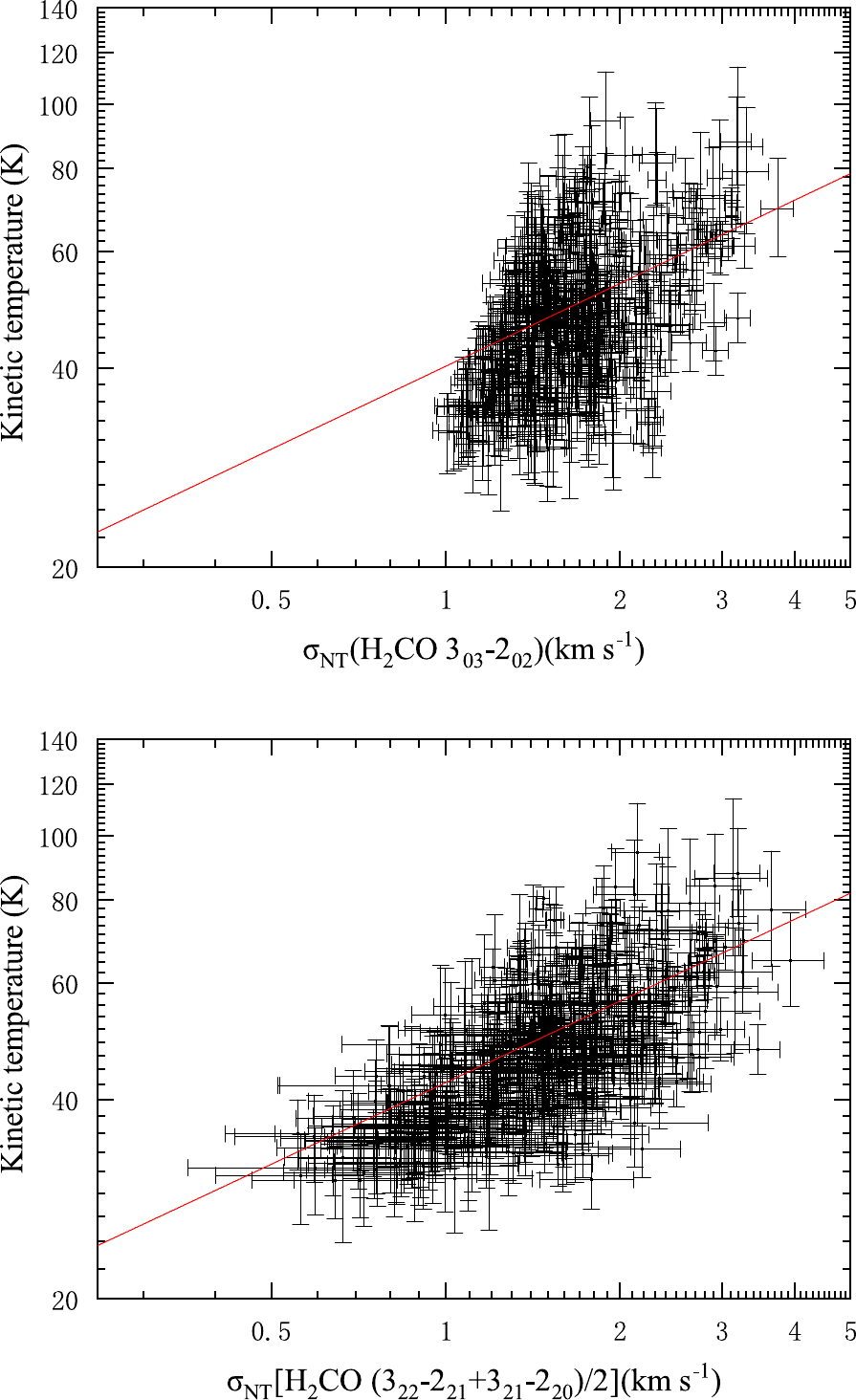}
    \caption{The relationship between the non-thermal linewidth ($\sigma_{\rm NT}$) and gas kinetic temperature, as determined
    from the H$_{2}$CO line ratio, is examined in those parts of the DR21 filament, which are characterized by a Mach number
    $\mathcal{M}$\,$\gtrsim$\,3.0 (see Sect.\,\ref{Sec:Turbulence}). A linear fit is applied to the H$_{2}$CO\,3$_{03}$--2$_{02}$
    and the combined H$_{2}$CO\,3$_{22}$--2$_{21}$ and 3$_{21}$--2$_{20}$ lines, yielding the following results:
    log$T_{\rm kin}=(0.42\pm0.04)\times{\rm log}\sigma_{\rm NT}+(1.60\pm0.01)$ and
    log$T_{\rm kin}=(0.41\pm0.03)\times{\rm log}\sigma_{\rm NT}+(1.63\pm0.01)$, respectively.
    In this fitting, the linear fit was performed on the unweighted data using the least squares method.
    The Pearson correlation coefficients, $R$, for these fits (red lines) are 0.45 and 0.55, respectively.}
    \label{fig:NT-T}
    \end{center}
    \end{figure}

We investigate the potential relationship between turbulence and temperature on a scale of $\sim$0.1\,pc within the DR21 filament.
To assess turbulence, we utilize the non-thermal linewidth ($\sigma_{\rm NT}$) of H$_{2}$CO, which serves as a reliable indicator.
Additionally, we determine the gas kinetic temperature using the H$_{2}$CO line ratio. Specifically, we focus on positions near dense regions
that exhibit strong non-thermal motions ($\mathcal{M}$\,$\gtrsim$\,3.0). As stated in Sect\,\ref {sec:Th-non-th}, the dense gas in locations
characterized by higher Mach numbers ($\mathcal{M}$\,$\gtrsim$\,3.0) is significantly impacted by non-thermal motions. Therefore, for our
analysis, we specifically choose positions in close proximity to W75S\,FIR\,3, W75S\,FIR\,2, DR21(OH)S, DR21, and DR21W1.
In Fig.\,\ref{fig:NT-T}, we present the correlation between the
non-thermal linewidth and gas kinetic temperature for both the H$_{2}$CO\,3$_{03}$--2$_{02}$ line and the combined
H$_{2}$CO\,3$_{22}$--2$_{21}$ and 3$_{21}$--2$_{20}$ lines. A power law relationship is fitted, with
$T_{\rm kin}\propto\sigma_{\rm NT}^{0.42\pm0.04}$ for H$_{2}$CO\,3$_{03}$--2$_{02}$
and $T_{\rm kin}\propto\sigma_{\rm NT}^{0.41\pm0.03}$ for the combined H$_{2}$CO\,3$_{22}$--2$_{21}$ and 3$_{21}$--2$_{20}$ lines.
The correlation coefficients, $R$, are determined to be 0.45 and 0.55, respectively.
These results suggest that the elevated temperature observed in regions of strong non-thermal motion within the DR21 filament,
as indicated by H$_{2}$CO, can be attributed to turbulence occurring at a scale of $\sim$0.1\,pc, which aligns with
previous observational studies conducted on massive star-forming clumps (ATLASGAL sample, \citealt{Tang2018b}) and the Orion
molecular cloud 1 (OMC-1, \citealt{Tang2018a}), where similar temperature variations were observed on a scale ranging from approximately
0.06 to 2\,pc. One should note that this agreement pertains solely to the intercept and not the power law index (see Fig.\,\ref{fig:NT-T}).
The power law index fitted in the DR21 filament is approximately 0.41-0.42, which is lower than the previous observational results
($T_{\rm kin}\propto\sigma_{\rm NT}^{0.76-1.26}$) in the ATLASGAL sample and OMC-1 \citep{Tang2018a,Tang2018b}.
As previously mentioned, the dense gas in the DR21 filament is also influenced by various factors, such as outflows, shocks,
and/or radiation from regions of massive star formation in W75S, DR21(OH), and DR21. The variation in the power law index could be
attributed to the disparity in the contribution fractions of gas heating mechanisms to the gas temperature between the DR21 filament
and the ATLASGAL sample, as well as OMC-1.

Assuming that the dominant factor heating the gas is turbulence, we employ the methodology outlined by \cite{Tang2018a} in their Equation\,(2)
to compute the gas kinetic temperature, denoted as $T_{\rm turb}$, resulting from turbulence in the DR21 filament. Our calculations are based
on the following assumptions: a gas density of $n$(H$_2$)\,=\,10$^5$\,cm$^{-3}$, a typical velocity gradient of 1\,km\,s$^{-1}$\,pc$^{-1}$,
the non-thermal linewidth values of H$_{2}$CO provided in Table\,\ref{table:Clumps-Parameters}, a cloud size of $\sim$1\,pc,
and a dust temperature equal to the gas temperature derived from H$_{2}$CO. The computed $T_{\rm turb}$ values are presented
in Table\,\ref{table:Clumps-Parameters}, revealing a notable correlation between the derived $T_{\rm turb}$ value and the kinetic temperature
obtained from the H$_{2}$CO line ratio. These results suggest that turbulence plays a significant role in determining the dense gas kinetic
temperature on a scale of  $\sim$0.1\,pc in the DR21 filament.

\subsection{Comparison with the OMC-1 and N113}
\label{Sec:OMC-1}
The Orion molecular cloud 1 is a well-known target for studying the physical and chemical properties of molecular clouds and their
role in star formation. Its complex gas heating has been revealed by observations made with the APEX 12\,m telescope over a region of size
$\sim$1.1$\times$1.7\,pc$^{2}$, where the same H$_{2}$CO\,(3--2) lines were mapped on a scale of $\sim$0.06\,pc \citep{Tang2018a}.
The complex gas heating in the OMC-1 is likely due to multiple processes including star formation activity, radiation, and turbulence.
The N113 region, which is located in the Large Magellanic Cloud, is a well-studied site for massive star formation and shows strong molecular
line emission (e.g., \citealt{Tang2017a,Tang2021,Gong2023a}). The complex gas heating in N113 was also studied using an ALMA survey over a relatively
large area of N113 ($\sim$2.4$\times$6.1\,pc$^{2}$), where the transitions of the same H$_{2}$CO\,(3--2) lines were detected on a linear
scale of $\sim$0.4\,pc \citep{Tang2018b}. The gas heating process in N113 appears to be as intricate as that of its counterpart in OMC-1.

As discussed above, our observations in the DR21 filament ($\sim$1.4$\times$4.4\,pc$^{2}$) on a scale of $\sim$0.1\,pc also indicate
the presence of similar gas heating mechanisms as in OMC-1 and N113. In addition, we compare the detailed physical parameters of dense gas,
H$_{2}$CO line ratio, $T_{\rm kin}$, $\sigma_{\rm T}$, $\sigma_{\rm NT}$, $c_{\rm s}$, and $\mathcal{M}$, derived from H$_{2}$CO\,(3--2)
lines in the DR21 filament, OMC-1, and N113 in Table\,\ref{table:Parameters_compared}. The median values of these physical parameters in
the dense gas within the DR21 filament appear to be nearly identical to the results obtained in OMC-1 and N113 on a scale of
$\sim$0.1--0.4\,pc (see Table\,\ref{table:Parameters_compared}). It should be noted that the mean values of non-thermal linewidth
and Mach number in OMC-1 are marginally lower than those observed in DR21 and N113. This discrepancy may be attributed to the superior
resolution of OMC-1 (beam size $\sim$0.06\,pc) and/or the inherent turbulence fluctuations across various molecular clouds.
These similarities in physical parameters suggest that the processes
governing the dynamics and thermodynamics of the dense gas traced by H$_{2}$CO in different star formation regions may share common
underlying principles, despite variations in specific environmental conditions. The dense gas in star formation regions is influenced by
similar environmental factors, such as gravitational collapse, turbulent motions, and thermal processes, which contribute to the consistency
in these parameters across various regions and evolutionary stages. The physical conditions in star-forming regions appear to be shaped by
similar mechanisms, leading to the convergence of these parameters. This can include processes such as supersonic turbulence,
thermal pressure, and the interaction between magnetic fields and turbulent motions, all of which contribute to the observed similarities
in the physical parameters.

\section{Summary}
\label{Sec:Sumamry}
The kinetic temperature structure of the massive filament DR21 in the Cygnus X molecular cloud complex has been mapped using the H$_{2}$CO
218\,GHz line triplet observed with the IRAM 30\,m telescope. The main results are the following:
\begin{enumerate}
\item
The H$_2$CO\,3$_{03}$--2$_{02}$ emission exhibits an extensive distribution and provides a clear measurement of the dense ridge structure
of the DR21 filament.

\item
By employing the RADEX non-LTE model, we have determined the gas temperature through the modeling of the measured
H$_2$CO\,0.5$\times$[(3$_{22}$--2$_{21}$\,+\,3$_{21}$--2$_{20})$/3$_{03}$--2$_{02}$] line ratios. The derived gas kinetic temperatures
exhibit a range, varying from 24 to 114\,K, with an average of 48.3$\pm$0.5\,K at a spatial density of 10$^{5}$\,cm$^{-3}$.
In comparison to temperature measurements utilizing NH$_3$\,(1,1)/(2,2) and FIR wavelengths, the H$_2$CO lines trace distinctly higher
temperatures.

\item
The dense gas, traced by H$_{2}$CO, exhibiting kinetic temperatures in excess of 50\,K appears to be correlated with the dense clumps
found in various regions, namely W75S\,FIR\,3, DR21(OH), DR21(OH)W, DR21(OH)S, DR21, and DR21W1. On the other hand, the outskirts of the
DR21 filament show lower temperature distributions ($T_{\rm kin}$\,$<$\,50\,K).

\item
Four dense cores, N44, N46, N48, and N54, located in the DR21(OH), DR21, DR21(OH)S, and W75S\,FIR\,3 regions, respectively,
exhibit temperature gradients on a scale of approximately 0.1–0.3\,pc. This suggests that the warm dense gas traced by H$_{2}$CO is
influenced by internal star formation activity. Unlike the dense core N54, the temperature profiles of these dense cores could be
fitted with power-law indices ranging from $-$0.3 to $-$0.5, with a mean value of $\sim-$0.4, indicating that the warm dense gas probed
by H$_{2}$CO is heated by radiation emitted from internally embedded protostar(s) and/or stellar clusters.

\item
In the DR21 region, our temperature measurements towards the dense core N46, located at the putative position of
a supposed major explosive event, provide no direct evidence supporting this notion on a scale of $\sim$0.1\,pc.
Nevertheless, we find compelling evidence that the dense gas in the DR21W1 region is indeed heated
by shocks originating from the western DR21 flow.

\item
The non-thermal linewidths of H$_{2}$CO exhibit a correlation with the gas kinetic temperatures within the DR21 filament.
This suggests that higher temperatures, as traced by H$_{2}$CO, are associated with turbulence on a scale of $\sim$0.1\,pc.

\item
The physical parameters of the dense gas, as determined from H$_{2}$CO lines in the DR21 filament, exhibit a striking similarity to the
results obtained in OMC-1 and N113 on a scale of approximately 0.1-0.4\,pc. This may imply that the mechanisms governing the dynamics
and thermodynamics of dense gas traced by H$_{2}$CO in diverse star formation regions may be governed by common underlying principles,
despite variations in specific environmental conditions.

\end{enumerate}

\begin{acknowledgements}
We thank the staff of the IRAM telescope for their assistance in observations.
This work acknowledges the support of the National Key R\&D Program of China under grant No.\,2023YFA1608002, the Chinese Academy of
Sciences (CAS) “Light of West China” Program under grant No.\,xbzg-zdsys-202212, the Tianshan Talent Program of Xinjiang Uygur Autonomous
Region under grant No.\,2022TSYCLJ0005, and the Natural Science Foundation of Xinjiang Uygur Autonomous Region under grant No.\,2022D01E06.
It was also partially supported by the National Key R\&D Program of China under grant No.\,2022YFA1603103, the CAS “Light of West China”
Program under grant No.\,2020-XBQNXZ-017, the National Natural Science Foundation of China under grant Nos.\,12173075 and 12373029,
and the Youth Innovation Promotion Association CAS. T.\,Liu, K.\,Wang, X.\,P.\,Chen, and J.\,W.\,Wu acknowledge
support by the Tianchi Talent Program of Xinjiang Uygur Autonomous Region. C.\,Henkel acknowledges support by the Chinese Academy of
Sciences President's International Fellowship Initiative under grant Nos.\,2023VMA0031 and 2025PVA0048.
This research has used NASA's Astrophysical Data System (ADS).
\end{acknowledgements}

\bibliographystyle{aa} 
\bibliography{bibfile} 

\begin{appendix} 
\onecolumn
\section{H$_2$CO Velocity Channel Maps}
\begin{figure*}[h]
\centering
\includegraphics[width=1.0\textwidth]{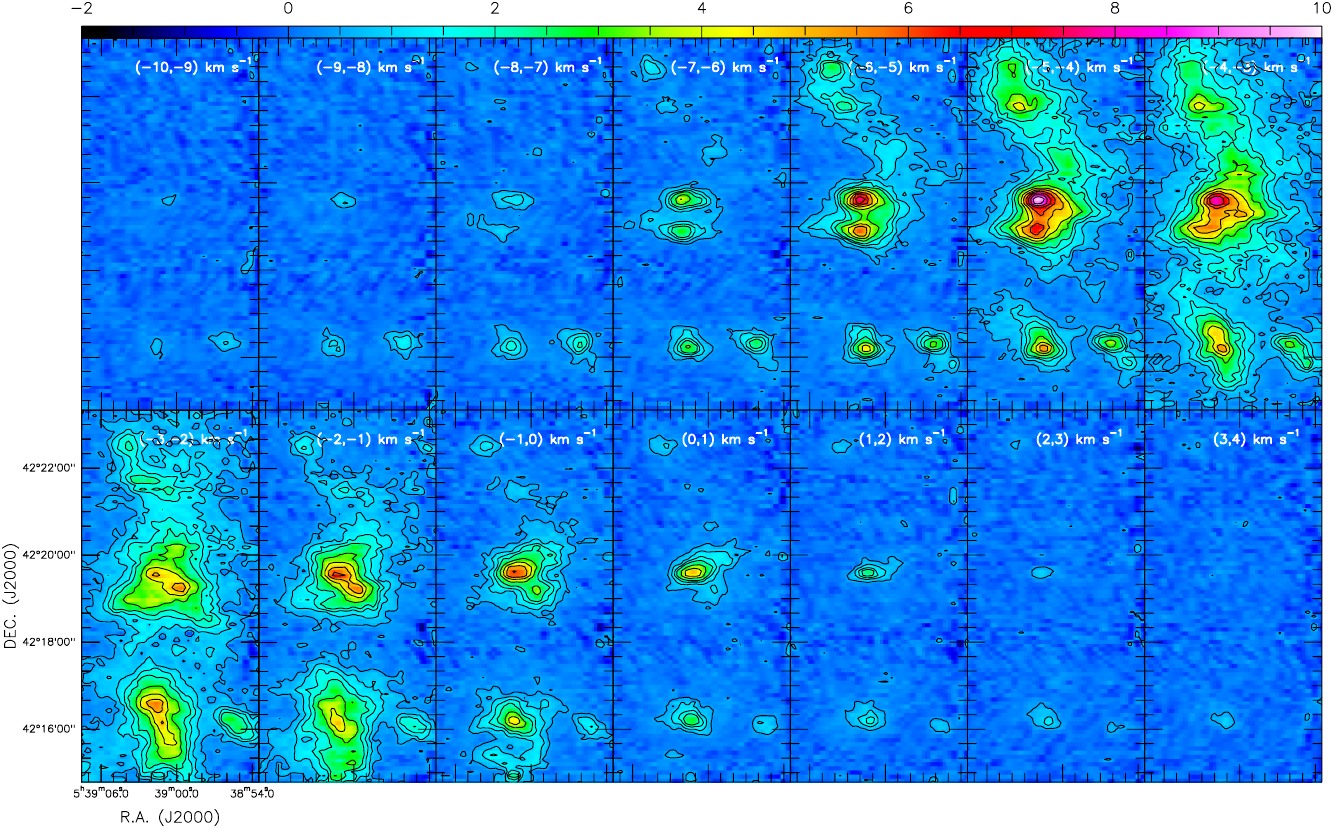}
\caption{Channel maps of the H$_{2}$CO\,(3$_{03}$--2$_{02}$) transition. The contour levels are running from 0.5 to 2.0\,K in steps of
0.5\,K and from 2 to 10\,K in steps of 1\,K\,($T_{\rm mb}$ scale; color bar in units of K).}
\label{fig:H2CO-channel}
\end{figure*}

\section{The LTE and Non-LTE Models for H$_2$CO}
\begin{figure}[h]
\vspace*{0.2mm}
\begin{center}
\includegraphics[width=0.43\textwidth]{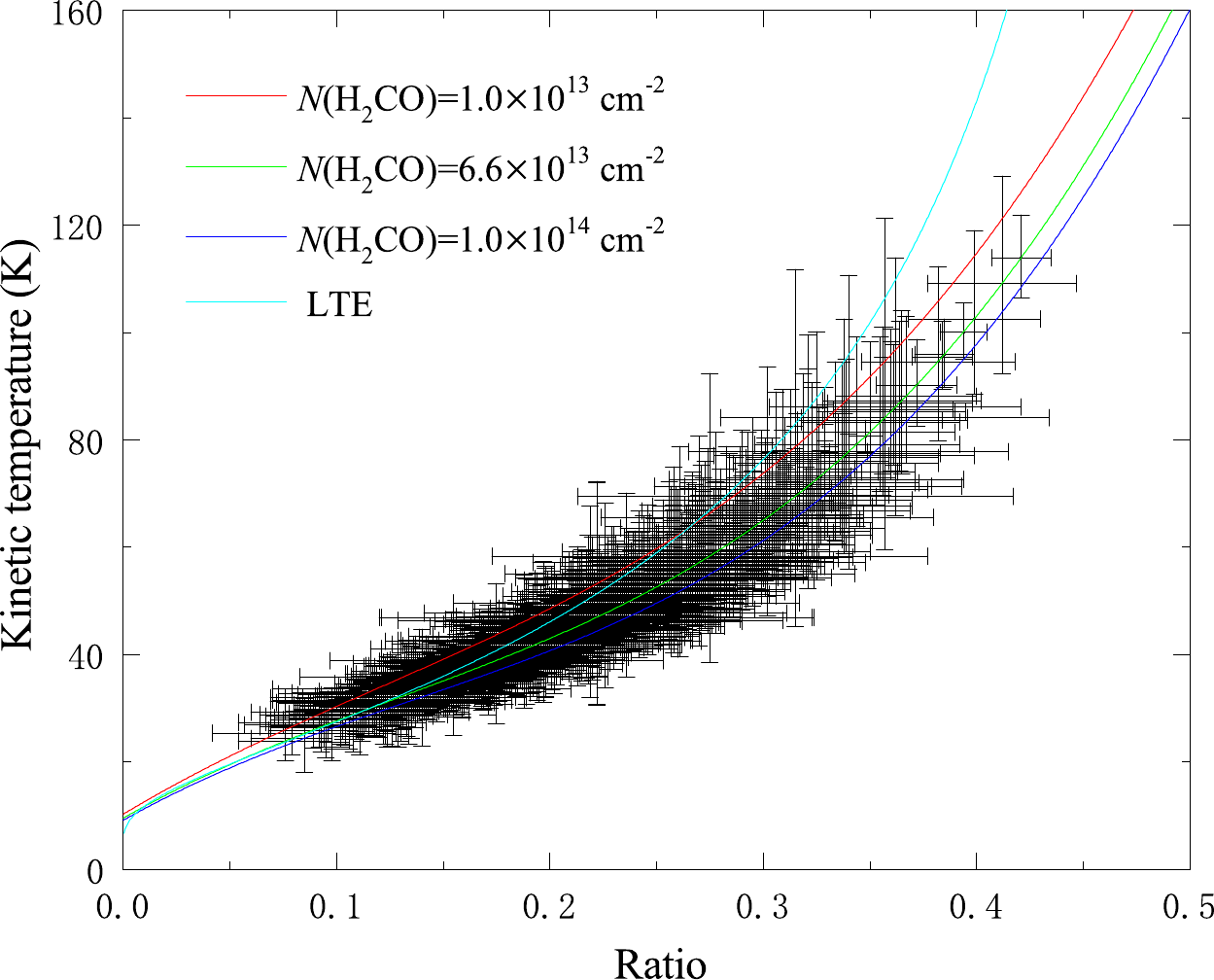}
\end{center}
\caption{RADEX non-LTE modeling of the relation between the kinetic temperature and the average ratio of
H$_{2}$CO\,3$_{22}$--2$_{21}$/3$_{03}$--2$_{02}$ and 3$_{21}$--2$_{20}$/3$_{03}$--2$_{02}$ with an assumed density of
$n$(H$_2$)\,=10$^{5}$\,cm$^{-3}$ and column densities $N$(para-H$_{2}$CO)\,=1.0$\times$10$^{13}$, 6.6$\times$10$^{13}$ and
1.0$\times$10$^{14}$\,cm$^{-2}$ (red, green, and blue), and the LTE (cyan), and an averaged linewidth of 3.8\,km\,s$^{-1}$.
The LTE kinetic temperature ($T_{\rm LTE}$) is determined using the methodology described in \cite{Mangum1993a} and \cite{Tang2017b}.
$T_{\rm LTE}$ is calculated as $T_{\rm LTE}\,=\,\frac{47.1}{{\rm ln}(0.556/{\rm Ratio})}~{\rm K}$, where Ratio represents
the average ratio of H$_{2}$CO\,3$_{22}$--2$_{21}$/3$_{03}$--2$_{02}$ and 3$_{21}$--2$_{20}$/3$_{03}$--2$_{02}$.
The black points are derived from our observed H$_{2}$CO line ratios from the DR21 filament for a column density
$N$(para-H$_{2}$CO)\,=\,6.6$\times$10$^{13}$\,cm$^{-2}$. The temperature uncertainties are obtained from observed H$_{2}$CO
line ratio errors.}
\label{fig:H2CO-radex}
\end{figure}

\end{appendix}

\end{document}